\documentclass[apjl]{emulateapj}
\usepackage{epsf}
\usepackage{amssymb,amsmath}
\usepackage{multirow}
\usepackage{rotating}
\usepackage{subfigure}
\usepackage{graphicx}
\usepackage{color}
\usepackage{epstopdf}

\newcommand\resetsubfigs{\setcounter{sub\@captype}{0}}
\newcommand{\appropto}{\mathrel{\vcenter{
  \offinterlineskip\halign{\hfil$##$\cr
    \propto\cr\noalign{\kern2pt}\sim\cr\noalign{\kern-2pt}}}}}

\def\mR{$-(0.22\pm0.06)$}
\def\bR{$-(1.7\pm0.4)$}
\def\sigR{0.35$^{+0.06}_{-0.05}$ }
\def\Prefl{$1.2\times10^{-4}$}
\def\PH{$3.7\times10^{-4}$}
\def\PHone{$1\times10^{-5}$}
\def\mH{ $0.25^{+0.07}_{-0.08}$  }
\def\bH{ $2.5\pm0.5$ }
\def\sigH{0.26$^{+0.11}_{-0.09}$}

\def\jetN{$-(10.1\pm0.4)$}
\def\jettheta{ $13^{+5}_{-4} {}^\circ$}
\def\jetLcrit{ $-(9.6\pm0.3)$ }
\def\jetL{39.3}
\def\jetsigR{0.46$^{+0.07}_{-0.06}$ }
\def\jetsigH{0.17$^{+0.12}_{-0.11}$ }
\def\jetV{$0.04< \beta <0.40$}

\def\jetDichRad{100} 

\begin{document}
\slugcomment{Accepted to ApJ on December 22, 2016}
\title{ AGN Coronae Through A Jet Perspective}
\author{Ashley L. King\altaffilmark{1,2}, Anne Lohfink\altaffilmark{3} \& Erin Kara\altaffilmark{4,5}}
\altaffiltext{1}{KIPAC, Stanford University, 452 Lomita Mall, Stanford, CA 94305 USA, ashking@stanford.edu}
\altaffiltext{2}{Einstein Fellow}
\altaffiltext{3}{Institute of Astronomy, Madingley Rd, Cambridge CB3 0HA}
\altaffiltext{4}{Department of Astronomy, University of Maryland, College Park, MD 20742-2421, USA}
\altaffiltext{5}{Hubble Fellow}

\begin{abstract}
This paper presents an in depth look at the jet and coronal properties of 41 AGN. Utilizing the highest quality {\it NuSTAR}, {\it XMM-Newton}, and NVSS 1.4 GHz data, we find that the radio Eddington luminosity inversely scales with X-ray reflection fraction, and positively scales with the distance between the corona and the reflected regions in the disk. We next investigate a model fit to the data that predicts the corona is outflowing and propagates into the large scale jet. We find this model describes the data well and predicts the corona has mildly relativistic velocities, \jetV. We discuss our results in the context of disk-jet connections in AGN. \end{abstract}

\keywords{}

\maketitle
\section{Introduction}
Understanding the symbiosis between accretion and outflows in the vicinity of black holes is vital to our understanding of how black holes impact their environments.  Outflows are ubiquitously associated with accretion disks and can liberate a great deal of energy through both radiation as well as through kinetic energy \citep[e.g][]{Begelman84,Allen06}. This feedback is thought to be responsible for large scale correlations between galaxy dynamics and the mass of the central supermassive black hole \citep[e.g. M$-\sigma$ relations, M$-$Luminosity relations][]{Greene06,Robertson06,Somerville08,Gultekin09b}, as well as heating of the central regions of clusters of galaxies, halting star-formation that should otherwise be forming from the cooling cluster medium \citep{McNamara07,Fabian12a,Prasad15}.

Highly collimated, relativistic jets are a particular type of outflow associated with accreting supermassive black holes and stellar-mass black holes in the X-ray spectrally ``hard-state'' \citep{Begelman84,Mirabel99}. Though unresolvable in stellar-mass black holes, the jets associated with supermassive black holes vary in size and structure, from compact cores less than a few parsecs \citep{Falcke00} to highly extended flows that reach scales of Megaparsecs \citep{Fanaroff74}. Furthermore, jets also vary in luminosity, typically by a factor of a few in supermassive black holes on yearly timescales \citep[e.g][]{Barvainis05,Chatterjee09} and a factor of a few decades in stellar-mass black holes on daily to monthly timescales \citep[e.g][]{Corbel00,Gallo03}. 

The changes in jet luminosity are ascribed to changes in the accretion flow \cite[e.g.][]{Heinz03, Merloni03}. Unfortunately, in stellar-mass black holes in the hard-state and in AGN, the thermal component directly probing the accretion disk is absent or heavily obscured by the host galaxy gas and dust \citep{Fender04,Martinez05,Merloni14}. This makes quantifying the changes in accretion rate difficult. Fortunately, the X-ray power-law that is readily observed is thought to probe the closest regions to the black holes \citep[e.g.,][]{Dai03,Pooley06,Dai10}.  Also known as the corona, this X-ray power-law is either generated via Comptonization of the disk or jet photons closest to the black hole \citep{Sunyaev80,Markoff01,Markoff05}, and is known to correlate well with jet luminosity in the other bands. 

The strong correlation between the X-ray corona and radio emission from the core of the jet was first observed in individual stellar-mass black hole sources like GX 339-4 \citep{Corbel00}, but later found to extend to all stellar-mass black holes in the hard-state \citep{Gallo03,Gallo12}. Contemporaneously, when the mass of the black hole was also added to describe the trend, researchers found that this relation also held true for supermassive black holes \citep{Merloni03,Falcke04,Kording06,Gultekin09,Nisbet16}. This plane became known as the ``fundamental plane of black hole activity'', and indicates that the innermost regions near the black hole (probed by X-rays) are connected to larger scale production of jets (probed by the radio band) in a ubiquitous way. However, there is debate as to whether the X-rays are formed from upscattered photons from the inner accretion flow or the base of the jet, both of which are located a few tens of gravitational radii from the black hole.

To address where the X-rays originate, \cite{Merloni03} use the measured coefficients from the fundamental plane of black hole activity to test various disk-jet models and assess the X-rays' origin. The authors find that in a statistical sense, they can rule out that the X-ray emission originates from a radiatively ``efficient'' accretion flow, and find that a radiatively inefficient accretion is their favored model. However, they can not rule out optically thin synchrotron emission from the base of a jet as the primary source of seed photons for the X-ray component. Conversely, \cite{Falcke04} preform a similar analysis and come to the conclusion that the X-ray emission is dominated by optically thin synchrotron emission from the jet. Followup work by \cite{Kording06} reconciles the two competing results by asserting that \cite{Merloni03} include a sample which spans a wider range in mass-accretion rates, whereas their own work focuses on very low-Eddington sources. \cite{Kording06} further demonstrate that the measured coefficients of the plane are extremely sensitive to the sample selection. Corroborating this idea that the planes examine sources with different accretion modes, \cite{Merloni03} find a suggestion of higher scatter above $\log L_X/L_{Edd} = -3$ and indicate this could be due to changes in the accretion states, similar to what is seen in the ``high-soft'' state of X-ray binaries. 

The suggestion that at very low mass-accretion rates, the X-ray luminosity from accreting black holes is arising from the jet, is also supported by other analyses. Strong correlations between optical/NIR and X-ray observations of black hole X-ray binaries indicate that the two bands are intimately connected, and likely both produced in the jet \citep{Homan05,Russell06}. Furthermore, conical jet models have been used to model broad-band spectral energy distributions of individual X-ray binaries, like GX 339-4, and can explain both the magnitudes and spectral slopes across the spectral distribution, assuming the emission is coming from the jet \citep[e.g.,][]{Markoff05}. These models also explain some of the curvature observed above 10 keV \citep{Markoff05}, and have even been extended to broad-band spectral energy distribution modeling in low-Eddington supermassive black holes, like M81*, NGC 4051, and M87 \citep{Markoff08,Maitra11,Prieto16}. 

Interestingly, these jet models assume that the base of the jet is located within a few tens of gravitational radii and are accelerated along a region of 100-1000 $R_G$, where $R_G=GM/c^2$ \citep{Markoff04,Markoff05}. As the jet region is assumed to be the source of the power-law X-ray component that shines down onto the accretion disk, a predicted relativistic velocity should have an impact on the reflection features that are generated and observed in the X-ray band \citep{Beloborodov99,Malzac01}. \cite{Markoff04} investigate this impact, and find that the mildly relativistic flow from their model predicts a reflection fraction of $\sim$10\%, consistent with the reflection fraction observed in GX 339-4.  The authors find that the velocity of the base of the jet is $\sim 0.3-0.4c$ ,and the X-ray emission is dominated by synchrotron self-Compton at the base of the jet in their model. Similar velocities for the corona are also found in the moving corona model by \cite{Beloborodov99}, which is also used to describe the X-ray binary, Cygnus X-1. These authors primarily focus on spectral features in the X-ray band, and did not include subsequent emission in the outflow at other wavelengths like the model presented by \cite{Markoff05}. 

The advantages of the model put forward by \cite{Beloborodov99} is that it is not limited to low Eddington sources like the model of \cite{Markoff04,Markoff05}. It does not assume anything about the accretion flow in the model \citep{Beloborodov99}. In addition, the model presented by \cite{Beloborodov99} and further explored via Monte-Carlo simulations by \cite{Malzac01}, not only characterizes the magnitude of the reflection fraction in X-ray binaries, but also characterizes the observed relation between reflection fraction and the X-ray power-law spectral index in both X-ray binaries and AGN \citep{Zdziarski99,Gilfanov99}. This correlation arises in the moving corona model because a corona that has a faster velocity away from the disk beams radiation away, which lowers the reflection fraction. At the same time, less emission is returning to the corona, reducing the amount of cooling in the corona, which creates a harder spectral index. Contrary to the models presented by \cite{Markoff04} and \cite{Markoff05}, the spectral index-reflection fraction correlation suggests that though the corona is outflowing and possibly the base of a jet, the seed photons arise from the disk rather than the base of the jet \citep{Beloborodov99,Malzac01}. Reconciling these two models is difficult, but the works may describe different accretion regimes. The sample of AGN in \cite{Zdziarski99} is dominated by Seyferts, which are generally thought to be at a relatively high Eddington fraction as compared to the X-ray binaries. In such a regime, one could expect disk photons to dominate over synchrotron or synchrotron self-Compton photons from the jet, and therefore may not be in strong contradiction with the low Eddington models put forth by \citep{Markoff04,Markoff05}. 

In the models presented by \cite{Beloborodov99}, \cite{Malzac01}, \cite{Markoff04}, and \cite{Markoff05}, the velocity of the corona is the key parameter in determining the reflection strength and spectral properties observed in accreting black holes.  Although not assessed quantitatively, this finding that a faster velocity results in a lower reflection fraction qualitatively seems to agree with observations of radio-quiet and radio-loud AGN. This is assuming the coronal velocities propagate into the jet, causing an increase of jet power when the corona velocity is higher. Radio-quiet Seyferts have strong reflection signatures and comparatively low jet powers compared to radio-loud Seyferts that have weak reflection features and strong jet powers \citep{Zdziarski95,Wozniak98,Eracleous00}. \cite{Wozniak98} even hint at this idea, as they suggest the small reflection fractions in their radio-loud AGN sample could be due to the continuum emission ``collimated'' away from the disk at mildly relativistic speeds, i.e. a jet. 

In this paper, we aim to expand upon these higher-Eddington AGN reflection studies with direct comparison to jet luminosities in order to further explore the nature of the X-ray corona. By quantitatively comparing the reflection features to the jet power, we will examine the viability of the model which postulates the corona is moving mildly relativistically at the base of a jet. 

\section{The AGN Sample}
Two AGN samples are included in this work. The first is comprised of 32 AGN observed with {\it NuSTAR} \citep{Harrison13}, which focuses on the reflection features observed in the broad-band X-ray spectra. The second is a sample of 22 AGN observed with {\it XMM-Newton} \citep{Jansen01}, which focuses on the X-ray reflection timing properties of these AGN. 13 AGN overlap between the two samples, see Table \ref{Tab:data}.
\begin{deluxetable*}{lllllllllll} 
\tabletypesize{\scriptsize}
\tablecolumns{11} 
\tablewidth{0pc} 
\tablecaption{AGN Sample} 
\tablehead{ 
\colhead{Name}  & \colhead{redshift}& \colhead{$\log M$ ($M_\odot$)} & Ref & \colhead{$\log(L_{1.4GHz}/L_{Edd})$ } & Ref & \colhead{$\theta_\mu \ (^\circ)$} & Ref  & \colhead{$\log R$ } & \colhead{$\log H$} ($R_G$)}  \\
\startdata

1H 0419-577 &  0.1040  &   8.00 $\pm 0.30$ &  $L_B-24$  &  -6.47$\pm 0.43$ &  1  & 53  &  40 &-0.40$^{+ 0.11}_{- 0.13}$ &   - \\ 
1H 0707-495 &  0.0406  &   6.37 $\pm 0.60$ &  $L_B-10$  &   - & -  & 69  &  49 & 0.75$^{+ 0.25}_{- 0.45}{}^*$ &    0.6$\pm  0.6$ \\ 
3C 120 &  0.0330  &   7.75 $\pm 0.04$ &  $r-8$  &  -4.81$\pm 0.31$ &  1  & 25  &  14 &-0.56$\pm 0.04$ &   - \\ 
3C 273 &  0.1583  &   8.40 $^{+ 0.08}_{- 0.11}$ &  $r-8$  &  -2.84$^{+ 0.32}_{- 0.33}$ &  1  & 6  &  32 &-1.05$^{+ 0.05}_{- 0.06}$ &   - \\ 
3C 382 &  0.0579  &   9.06 $\pm 0.42$ &  $L_B-13$  &  -5.83$\pm 0.52$ &  1  & 45  &  34 &-0.78$^{+ 0.06}_{- 0.07}$ &   - \\ 
3C 390.3 &  0.0562  &   8.64 $^{+ 0.04}_{- 0.05}$ &  $r-8$  &  -5.06$\pm 0.31$ &  1  & 27  &  32 &-0.69$^{+ 0.05}_{- 0.06}$ &   - \\ 
4C 74.26 &  0.1040  &   9.62 $\pm 0.60$ &  $L_B-10$  &  -6.87$\pm 0.67$ &  1  & 43  &  36 &-0.30$\pm 0.04$ &   - \\ 
Ark 120 &  0.0327  &   8.07 $^{+ 0.05}_{- 0.06}$ &  $r-8$  &  -7.58$\pm 0.31$ &  1  & 45  &  45 &-0.21$\pm 0.04$ &   - \\ 
Ark 564 &  0.0247  &   6.00 $\pm 0.30$ &  $r-8$  &  -5.36$\pm 0.43$ &  1  & 45  &  52 &-0.29$\pm 0.05$ &    1.3$^{+  0.5}_{-  0.8}$ \\ 
Cen A &  0.0018  &   8.38 $^{+ 0.40}_{- 0.54}$ &  $g-9$  &  -6.08$^{+ 0.50}_{- 0.62}$ &  3  & 76  &  35 &-0.57$^{+ 0.02}_{- 0.03}{}^n$ &   - \\ 
Cyg A &  0.0561  &   9.46 $^{+ 0.09}_{- 0.12}$ &  $g-9$  &  -3.72$^{+ 0.32}_{- 0.33}$ &  1  & 80  &  32 & $<-1.21{}^*$ &   - \\ 
ESO 362-G18 &  0.0124  &   7.65 $^{+ 0.12}_{- 0.17}$ &  $r-8$  &  -7.95$^{+ 0.33}_{- 0.35}$ &  1  & 53  &  59 & - &   0.8$\pm  0.3$ \\ 
Fairall 9 &  0.0470  &   8.30 $\pm 0.10$ &  $r-8$  &   - & -  & 11  &  48 &-0.33$\pm 0.04$ &   - \\ 
IC 4329A &  0.0161  &   7.00 $\pm 0.10$ &  $r-8$  &  -6.41$\pm 0.32$ &  1  & 18  &  33 &-0.52$\pm 0.02$ &    1.1$^{+  0.3}_{-  0.4}$ \\ 
IRAS 13224-3809 &  0.0658  &   6.80 $\pm 0.50$ &  $L_B-19$  &  -5.98$\pm 0.59$ &  1  & 60  &  52 & 0.36$^{+ 0.12}_{- 0.06}{}^*$ &    1.0$^{+  0.7}_{-  0.8}$ \\ 
IRAS 17020+4544 &  0.0604  &   6.54 $\pm 0.50$ &  $L_B-19$  &  -4.51$\pm 0.59$ &  1  & 8  &  56 & - &   0.9$^{+  0.7}_{-  1.0}$ \\ 
MCG -05-23-16 &  0.0085  &   7.60 $\pm 0.70$ &  $OIII-16$  &  -8.25$\pm 0.76$ &  1  & 38  &  44 &-0.49$\pm 0.02$ &    0.7$^{+  0.9}_{-  1.0}$ \\ 
MCG -06-30-15 &  0.0077  &   6.46 $^{+ 0.21}_{- 0.35}$ &  $X-18$  &  -8.10$^{+ 0.37}_{- 0.46}$ &  4  & 33  &  46 & 0.01$^{+ 0.03}_{- 0.02}$ &   - \\ 
MS 2254.9-3712 &  0.0390  &   6.60 $\pm 0.50$ &  $X-22$  &  -6.15$\pm 0.00$ &  1  & 39  &  60 & - &   1.9$^{+  0.7}_{-  0.9}$ \\ 
Mrk 335 &  0.0258  &   7.23 $\pm 0.04$ &  $r-8$  &  -7.18$\pm 0.31$ &  1  & 60  &  27 & 0.55$\pm 0.04$ &    0.4$^{+  0.2}_{-  0.3}$ \\ 
Mrk 509 &  0.0344  &   8.05 $\pm 0.04$ &  $r-8$  &  -7.32$\pm 0.31$ &  1  & 56  &  51 &-0.32$^{+ 0.07}_{- 0.04}$ &   - \\ 
Mrk 766 &  0.0129  &   6.82 $^{+ 0.05}_{- 0.06}$ &  $r-8$  &  -6.66$\pm 0.31$ &  1  & 34  &  28 &-0.10$\pm 0.06$ &   - \\ 
NGC 1365 &  0.0055  &   7.66 $\pm 0.30$ &  $X-12$  &  -7.25$\pm 0.43$ &  1  & 65  &  31 & 0.20$\pm 0.03{}^*$ &    0.3$\pm  0.4$ \\ 
NGC 2110 &  0.0078  &   8.30 $\pm 0.30$ &  $\sigma-10$  &  -7.69$\pm 0.43$ &  1  & 72  &  38 &-0.74$^{+ 0.07}_{- 0.08}$ &   - \\ 
NGC 3516 &  0.0088  &   7.40 $^{+ 0.04}_{- 0.07}$ &  $r-8$  &  -7.66$\pm 0.31$ &  1  & 38  &  43 &-0.02$\pm 0.03$ &   - \\ 
NGC 3783 &  0.0097  &   7.40 $\pm 0.08$ &  $r-8$  &  -7.42$\pm 0.32$ &  1  & 22  &  57 & - &   0.1$^{+  0.2}_{-  0.3}$ \\ 
NGC 4051 &  0.0023  &   6.13 $^{+ 0.10}_{- 0.20}$ &  $r-8$  &  -7.08$^{+ 0.32}_{- 0.37}$ &  1  & 22  &  30 & 0.17$\pm 0.03$ &    1.1$\pm  0.3$ \\ 
NGC 4151 &  0.0033  &   7.56 $\pm 0.05$ &  $r-8$  &  -7.62$\pm 0.31$ &  1  & 18  &  25 & 0.00$^{+ 0.04}_{- 0.03}$ &    0.7$^{+  0.2}_{-  0.3}$ \\ 
NGC 4593 &  0.0090  &   6.88 $^{+ 0.08}_{- 0.10}$ &  $r-8$  &  -7.93$\pm 0.32$ &  1  & 32  &  50 &-0.26$\pm 0.04$ &   - \\ 
NGC 5506 &  0.0062  &   7.40 $\pm 0.30$ &  $\sigma-11$  &  -6.93$\pm 0.43$ &  1  & 44  &  37 &-0.17$\pm 0.03$ &    0.5$^{+  0.5}_{-  0.7}$ \\ 
NGC 5548 &  0.0172  &   7.12 $\pm 0.02$ &  $r-8$  &  -6.84$\pm 0.31$ &  1  & 46  &  26 &-0.43$\pm 0.02$ &    0.7$\pm  0.2$ \\ 
NGC 6860 &  0.0149  &   7.60 $\pm 0.50$ &  $L_B-19$  &  -7.81$\pm 0.59$ &  7  & 61  &  58 & - &   0.3$^{+  0.7}_{-  0.9}$ \\ 
NGC 7213 &  0.0058  &   7.99 $\pm 0.30$ &  $\sigma-10$  &  -7.33$\pm 0.43$ &  5  & 21  &  39 &-0.52$\pm 0.05$ &   - \\ 
NGC 7314 &  0.0048  &   6.78 $\pm 0.30$ &  $g-20$  &  -7.54$\pm 0.43$ &  1  & 9  &  39 &  - &    0.4$^{+  0.4}_{-  0.5}$ \\ 
NGC 7469 &  0.0163  &   6.96 $^{+ 0.05}_{- 0.05}$ &  $r-8$  &  -5.92$\pm 0.31$ &  1  & 23  &  55 & - &   1.6$^{+  0.3}_{-  0.7}$ \\ 
PDS 456 &  0.1840  &   9.00 $^{+ 0.50}_{- 1.00}$ &  $X-17$  &  -6.69$^{+ 0.59}_{- 1.05}$ &  1  & 70  &  40 & 0.11$^{+ 0.14}_{- 0.07}$ &   - \\ 
PG 1211+143 &  0.0809  &   8.16 $\pm 0.13$ &  $r-8$  &  -7.60$\pm 0.33$ &  6  & 44  &  29 & 0.45$^{+ 0.06}_{- 0.07}$ &    0.2$^{+  0.4}_{-  0.9}$ \\ 
PG 1244+026 &  0.0482  &   7.26 $\pm 0.50$ &  $\sigma-21$  &  -6.70$\pm 0.59$ &  1  & 38  &  53 &  - &    0.9$\pm  0.7$ \\ 
PG 1247+267 &  2.0380  &   8.92 $^{+ 0.15}_{- 0.17}$ &  $r-8$  &  -6.10$^{+ 0.34}_{- 0.35}$ &  2  & 34  &  41 & $<0.95$ &   - \\ 
REJ 1034+396 &  0.0424  &   6.30 $\pm 0.30$ &  $X-23$  &  -5.26$\pm 0.43$ &  1  & 30  &  54 & - &   1.7$^{+  0.5}_{-  0.6}$ \\ 
SWIFT J2127.4+5654 &  0.0144  &   7.18 $\pm 0.60$ &  $L_B-14$  &  -7.70$\pm 0.67$ &  1  & 49  &  47 & 0.23$\pm 0.03$ &    0.7$^{+  0.7}_{-  0.8}$ \\ 
\enddata
\tablecomments{ This table lists all the AGN utilized in this analysis. The reflection fraction, $\log R$, is determined via broad band spectral fits to {\it NuSTAR} data, while the distance between the corona and emitting regions in the disk, $\log H$, is determined via X-ray reverberation lags in {\it XMM-Newton} data \citep{Kara15}. The time lags do not include the additional dilution factor that is later applied during the analysis.  ${}^{*}$ denotes the fits which include an additional broad Fe K$\alpha$ line via \texttt{relline} \citep{Dauser15}.  ${}^{n}$ denotes the fits that include additional absorption in the broad band X-ray spectral fits. References: 1 \cite{Condon98}, 2 extrapolated from 4.9 GHz \cite{Barvainis96}, 3 \cite{Condon96}, 4 \cite{Ulvestad84}, 5 \cite{Wright94}, 6 \cite{Becker95}, 7 \cite{Mauch03}, 8 reverberation masses \cite{Bentz15}, 9 gas dynamics masses \cite{Marconi03}, 10 stellar-velocity dispersion, M-$L_B$ \cite{Woo02}, 11 stellar-velocity dispersion \cite{Papadakis04}, 12 X-ray scaling \cite{Kara15}, 13 M-$L_B$ \cite{Marchesini04}, 14 M-$L_B$ \cite{Malizia08}, 15 X-ray scaling \cite{Iwasawa16},16 O III FWHM \cite{Wandel86}, 17 X-ray scaling \cite{Nardini15}, 18 X-ray scaling \cite{McHardy05}, 19 M-$L_B$ \cite{Ponti12}, 20 gas dynamics \cite{Schulz94}, 21  velocity dispersion \cite{Marconi08}, 22 X-ray scaling \cite{Alston15}, 23 X-ray scaling \cite{Alston14}, 24 M-$L_B$ \cite{Fabian05}, 25 \cite{Keck15}, 26 \cite{Steenbrugge03}, 27 \cite{Gallo15},  28 \cite{Brenneman09}, 29 \cite{Lobban16}, 30 \cite{Ballantyne01}, 31 \cite{Walton14}, 32 \cite{Jorstad05}, 33 \cite{McKernan04}, 34 \cite{Giovannini01},  35 \cite{Furst16}, 36 \cite{Hasenkopf02}, 37 \cite{Matt15}, 38 \cite{Rosario10}, 39 \cite{Ruschel14}, 40 \cite{Walton10},  41 \cite{Lanzuisi16}, 42 \cite{Guainazzi16}, 43  \cite{Noda16}, 44 \cite{Zoghbi13} , 45 \cite{Garcia14}, 46 \cite{Marinucci14}, 47 \cite{Marinucci14b}, 48 \cite{Lohfink16},  49 \cite{Fabian12}, 50 \cite{Guainazzi99}, 51 \cite{Boissay14},  52 \cite{Chainakun16}, 53 \cite{Kara14},  54 \cite{Czerny10}, 55 \cite{Patrick11}, 56 \cite{Leighly99}, 57 \cite{Brenneman11}, 58 \cite{Bennert06}, 59 \cite{Agis14}, 60 no inclination was found; we therefore used the average viewing angle from the timing sample of AGN. \label{Tab:data}}
\end{deluxetable*}

\subsection{The {\it NuSTAR} Reflection Fraction Sample}
{\it NuSTAR} is a powerful satellite that measures the X-ray spectra between 3--79 keV \citep{Harrison13}. This band is extremely well suited for studying accretion onto black holes, as this is the band where we expect to observe distinct features from the accretion disk.  The most prominent features are the relativistically broadened Fe K$\alpha$ line at 6.4--6.97 keV, and the Compton hump at $\sim20$ keV \citep[e.g.,][]{George91,Reynolds99}. In addition, both of these features probe the corona, as a distribution of high energy particles above the disk is required to shine down and fluoresce the Fe K$\alpha$ line as well as scatter photons into the Compton hump spectral feature. 

Thirty two AGN are publicly available in the {\it NuSTAR} archive as of 1 August 2016 that meet the criterion for this study. This criterion includes Compton-thin AGN, i.e. $N_H<10^{24}$ cm$^{-2}$, with at least 20 ks exposure time. The average total exposure time of $\langle t_{exp}\rangle= 160 ks$, with an average total counts of $\langle Cts\rangle=1.8\times10^5$ photons.

Each of these AGN  {\it NuSTAR} data were reduced using the standard reduction pipeline {\it nupipeline}, during which the screening \texttt{saamode} was set to strict to ensure a low background level.  We produced {\it NuSTAR} spectra from the cleaned event files using the tool \texttt{nuproducts}. A 60 arcsec and 120 arcsec circular extraction region was used for the source and background regions, respectively. Both focal plane modules (FPM) A and B spectra were produced. For the targets with multiple observations, the spectra of each focal plane module were summed to create average spectra. The AGN are expected to be variable in the X-ray band, both in flux and spectral features \citep[e.g.][]{Nandra00,Fabian12,Parker14,Wilkins14,Wilkins15}. However, as we are comparing the reflection measurements to radio measurements taken years apart from the X-ray observations, variability between the observations likely dominates the systematic errors. In addition, because dynamical timescales scale with mass of the black hole, the X-ray variability will primarily effect our least massive sources. We therefore concentrate on the average X-ray properties of the AGN and include extra uncertainties in the radio band to account for variability between the measurements (See Section 2.3). 

The average FPMA and FPMB spectra were re-binned to include 20 counts per bin for the spectral modeling. The 3-78 keV energy range was used for the spectral modeling. We fit the spectra in ISIS \citep{Houck00} with a model based on a phenomenological cutoff power-law continuum. We limit the high energy cutoff of the power law to be larger than 75\,keV to avoid confusion with the Compton hump. The \texttt{xillver} model \citep{Garcia13} is then used to fit the Fe K$\alpha$ line and Compton hump. It simply and uniformly quantifies the total reflection fraction, which is measured as $R=\Omega/2\pi$, where $\Omega$ is the solid angle that the corona subtends. In most cases, the resolution of NuSTAR does not allow for disentangling the neutral from ionized reflection components, and therefore we only include one \texttt{xillver} component to quantify both. See Section 4.1 for further discussion.

The inclination parameter included in \texttt{xillver} is fixed at the value given in Table \ref{Tab:data} because it cannot be precisely constrained from our data. These values were primarily selected from higher resolution reflection fits which can measure the inclination of the inner disk \citep[e.g.,][]{Walton14,Garcia14,Kara14}. When this was not available, we utilized inclinations from the jet, assuming the jet axis is aligned perpendicular to the disk \citep[e.g.,][]{Giovannini01,Jorstad05}. Finally, we used inclinations of the infrared torus when no other estimates were available \citep{Ruschel14}. The average viewing inclination for our sample is $\langle \theta_{inc} \rangle = 41^\circ$, which is to be expected as our sample selects against type II AGN that are viewed close to edge on, i.e. large inclinations. 

To model the Fe K$\alpha$ line well, the iron abundance is allowed to vary from 0.5 to 3 times solar, which incorporates the typical values measured for most AGN \citep{Walton13}. The whole model is modified by the absorption from our own Galaxy described by the model \texttt{TBnew} with the cross section set to \texttt{vern} \citep{Verner96} and the abundances set to \texttt{wilm} \citep{Wilms00}. The absorption column is kept fixed to the average value from the $N_\mathrm{H}$ Tool for each source. This very basic baseline model provides a good fit ($\chi^2_\text{red}<1.5$) to 25 of the 30 sources. For the five sources with reduced $\chi^2$ larger than 1.5, we included excess absorption and/or a board Fe K$\alpha$ line \citep[\texttt{relline},][]{Dauser15}. These components are denoted in Table \ref{Tab:data}. The 1$\sigma$ errors for the reflection fraction are also given in Table \ref{Tab:data}, and where the reflection fraction is only an upper limit, the 3$\sigma$ errors are given.

Finally, in several of the radio-loud sources, one might expect an additional jet component in the continuum that would artificially dilute the reflection signal. In our sample, the strongest radio sources are Cyg A, 3C 382, Cen A, 3C 390.3, 3C 273, 4C 74.26, and 3C 120. However, in individual detailed analysis of these sources, we find our reflection fractions are consistent with those published in the literature using the same {\it NuSTAR} data sets, albeit sometimes slightly lower \citep[e.g.,][]{Madsen15,Reynolds15,Furst16}. As theses studies test for the possibility of an additional jet component but do not find a significant contribution from one, we may be underestimating the reflection fraction in a few of these sources but not by a significant amount.

\subsection{The {\it XMM-Newton} Fe K$\alpha$ Time-Lag Sample}
{\it XMM-Newton} is also a powerful X-ray satellite whose higher collecting area and spectral resolution surpasses that of {\it NuSTAR}, but with a more limited band pass, 0.3--10 keV \citep{Jansen01}. With its high collecting area, it is well suited for timing analysis of even the faintest X-ray sources. X-ray reverberation features of the inner accretion disk are expected on dynamical timescales, which for supermassive black holes occur on order of a few hundred seconds. 

In particular, one of the most prominent timing features is the high-frequency lag associated with the Fe K$\alpha$ region lagging behind the X-ray continuum. The size and shape of the time lags are quite informative, as they encode information about the height and geometry of both the emitting and reflecting regions  \citep{Wilkins16}. We assume the amplitude of the time lag, $t_{lag}$, is a good estimate of the extra light path taken by the photons that get reflected off the accretion disc,  $H=\frac{t_{lag}c^3}{GM}$. However, this does not include the effect of dilution on the lag \citep[see ][for detailed discussions]{Kara16,Uttley14}. In short, the time lag is measured between energy bands that contain both primary and reflected emission, and therefore this dilutes the actual light travel reverberation lag. This dilution effect needs to be modeled individually for each source \citep[e.g.][]{Cackett14}, and therefore the uncorrected lags presented in Table \ref{Tab:data} should be taken as a lower limit on the light travel time. Dilution corrections may increase the reverberation lag, and we account for this in Section 3.2.3.

Our Fe K$\alpha$ time lag sample comes from \cite{Kara16}. The authors have undertaken a comprehensive X-ray spectral timing analysis of all Seyfert AGN in the {\it XMM-Newton} archive that have at least 40 ksec exposures and show substantial variability.  Of the 43 AGN analyzed, 21 sources have detected X-ray lag signatures. The amplitude of these time lags range from 50--1800 seconds. In addition, eleven of these sources overlap with the {\it NuSTAR} reflection sample, although the {\it XMM-Newton} observations are not contemporaneous with the {\it NuSTAR} observations (Table \ref{Tab:data}).

\subsection{Radio Luminosities}
As the X-ray band probes the coronal properties, the radio band probes the jet power via synchrotron radiation from the accelerated particles along the magnetic fields in the jet \citep{Allen06,Merloni07}. The radio flux densities are primarily taken from the NRAO VLA Sky Survey (NVSS) \citep{Condon98}. This is a 1.4 GHz survey covering the entire sky north of $-40^\circ$ declination. It has a restoring beam of 45 arcsec at full-width half-maximum, and reaches a root-mean-square of 0.45 mJy. Given a $5\sigma$ detection limit, the faintest sources in the catalog are $>2.3$ mJy. Several sources were either located outside the survey area or undetected. Whenever possible, we included measurements at 1.4 GHz, but a select few sources, PG 1247+267 and NGC 7213,  were extrapolated from either 843 MHz or 5 GHz. See Table \ref{Tab:data} for the radio luminosities.

The majority of the sources in our sample are point-like sources. However, several radio-loud sources (4C 74.26, 3C309.3, 3C 382, Cen A, Cyg A) do show extended emission even at the NVSS survey resolution. We restrict the measurements to the core component, though some emission from the lobes may have been included due to the large restoring beam.

The radio measurements are not simultaneous with the reflection or reverberation measurements. In fact, most measurements are made over a decade apart. Variability in the radio-quiet and radio-loud is generally on order of 20\% on monthly timescales \citep{Barvainis05}, and larger variations are expected on yearly timescales \citep[e.g.][]{Chatterjee09,Chatterjee11}. Therefore, to address this variability, we include a factor of 0.3 dex in the systematic uncertainty of the radio luminosities, which dominates over any measurement uncertainties. 

\begin{figure*}[!th]
\center
\subfigure[][]{\includegraphics[width=.45\linewidth]{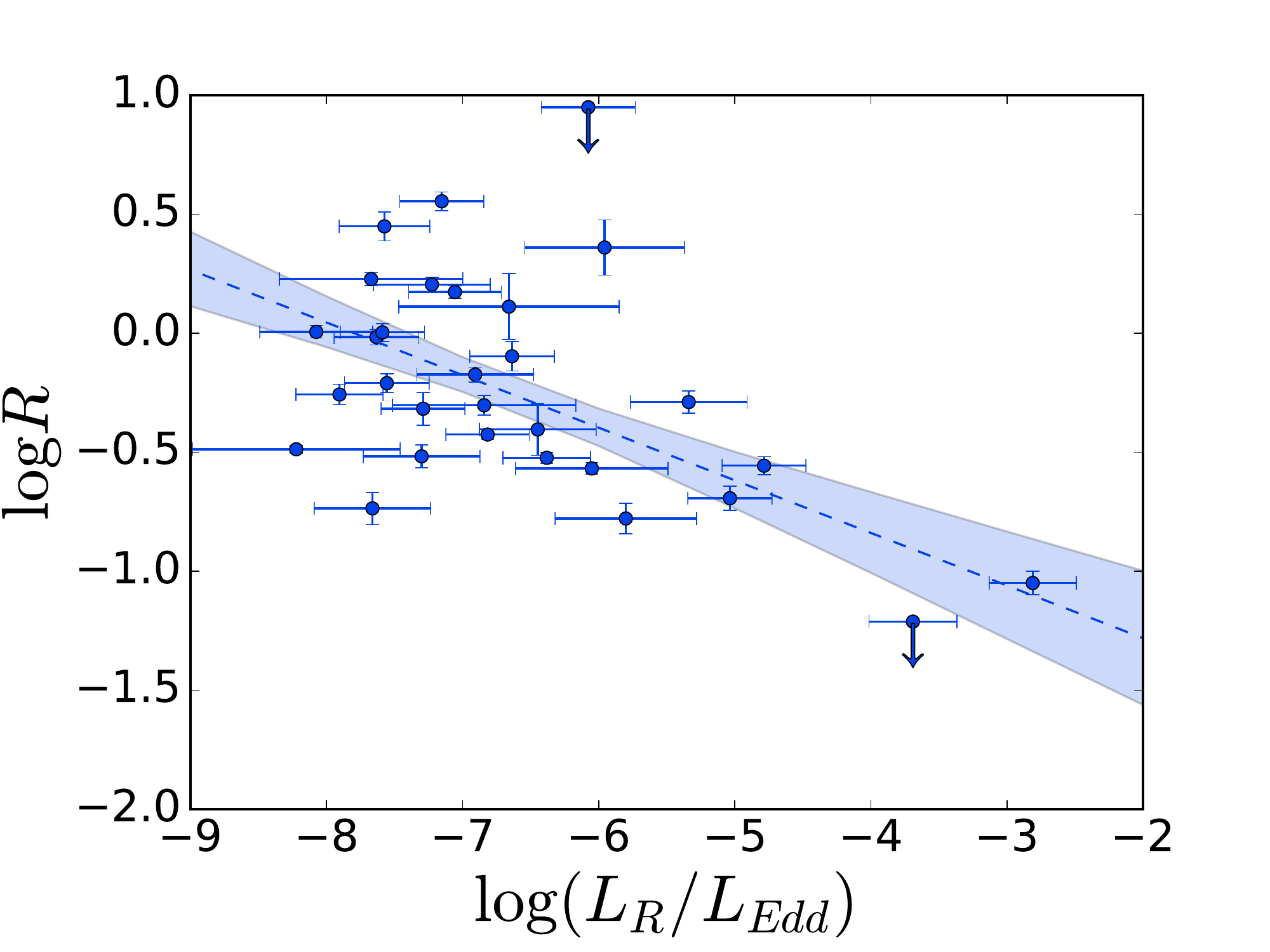}}
\subfigure[][]{\includegraphics[width=.45\linewidth]{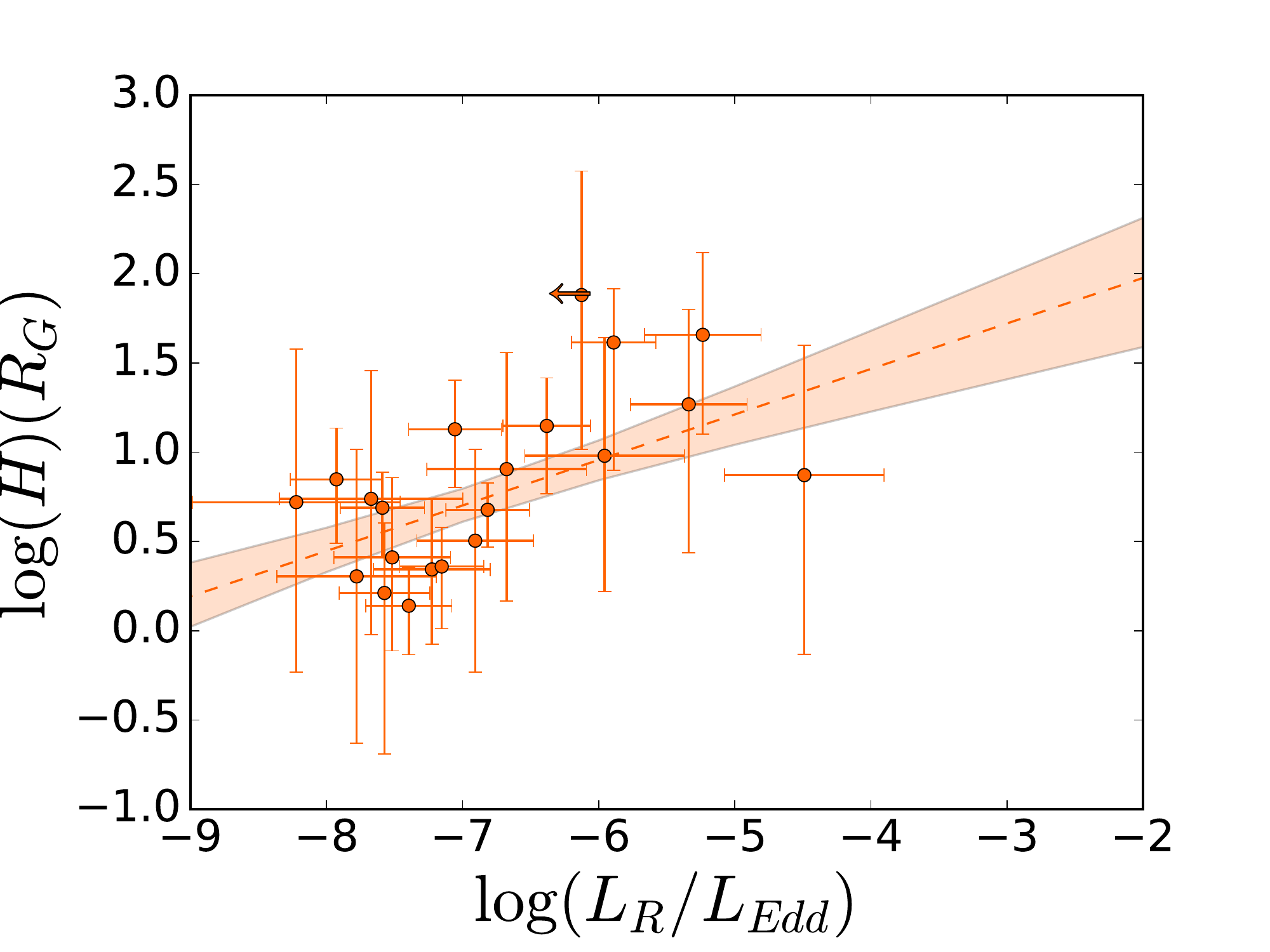}}
\caption{ Panel a) shows the {\it NuSTAR} reflection fraction versus the radio Eddington luminosity. There is an inverse correlation between these two parameters, and a slope of 0 can be ruled out with a probability of $p=$\Prefl. Panel b) shows {\it XMM-Newton} sample estimated path lengths between the corona and disk derived from reverberation lag measurements, versus the radio Eddington luminosity. A positive correlation is observed, and a slope of 0 is inconsistent at the $p=$\PH  confidence level. Both panels suggests changes in the corona as the radio power increases. Error bars are $1\sigma$.  \label{fig:data} }
\end{figure*}

\subsection{Black Hole Mass Estimates}

The radio luminosities are scaled by their black hole masses, which allows for self-similar comparison between the AGN. We refer to this radio luminosity as the ``radio Eddington luminosity'', which we explicitly define as the 1.4 GHz radio luminosity, $L_R$, divided by the Eddington luminosity, $L_{Edd}=1.38\times10^{38} (M)$ ergs $s^{-1}$, where $M$ is the mass of the supermassive black hole. The black hole masses are primarily derived from optical reverberation measurements, assuming a scaling factor $\langle f\rangle=4.3$ \citep{Bentz15,Grier13}. For those sources where reverberation masses are not available, gas kinematics, virial velocities, optical luminosity and X-ray scaling relations are all used in order to estimate the black hole masses. We list the masses in Table \ref{Tab:data}. The uncertainty for these mass estimates range from a few percent to 0.6 dex in the case of the X-ray scaling relations. PDS 456 has the most uncertain mass estimate as \cite{Nardini15} use the measured X-ray luminosity with an expected Eddington ratio to estimate its mass. 

The black hole mass distribution spans four decades in mass, with an average mass of $\langle \log M\rangle=7.6 M_\odot$, and a standard deviation of $\sigma$=0.9. The black holes with reflection measurements have an average mass of $\langle \log M\rangle= 7.8 M_\odot$ and $\sigma=0.9$ , while the black holes with X-ray reverberation measurements have a slightly smaller mass distribution at $\langle \log M\rangle=7.1 M_\odot$ and $\sigma=0.5$. The black holes are also accreting at relatively high mass-accretion rates $L_X/L_{Edd} \gtrsim 10^{-3}$, indicating the disk should be a thin disk rather than an advection dominated regime\citep[e.g.,][]{Narayan95,Narayan05}. This assumes the bolometric luminosity should be at least an order of magnitude higher than the X-ray luminosity.

\section{Analysis}
\subsection{Linear Fits}
Figure \ref{fig:data}a shows the reflection fraction versus radio Eddington luminosity. There is an apparent inverse correlation between the two parameters, though the sample is dominated by Seyfert galaxies clustered at low radio Eddington fractions. To test the strength of this correlation, we used a Markov chain Monte-Carlo  (MCMC) to fit a linear fit to the {\it NuSTAR} reflection fraction data in log space with an intrinsic scatter, implemented with python's {\it PyMC} v2.3.6 \citep{Fonnesbeck15}. The detected data points were drawn from a Normal distribution determined by their 1-$\sigma$ error bars. The data points with upper limits were drawn from a uniform distribution in linear space between zero and the 3-$\sigma$ upper limit. The fits were only marginally dependent on the upper limit priors.  We used 100,000 steps and discarded the initial third of the distribution.

The resulting best-fit relation is  $\log R =$\mR$\log(L_R/L_{Edd})$\bR and an intrinsic scatter of $\sigma_R$=\sigR (Table \ref{Tab:modelparams}). We test this against our null-hypothesis of no correlation, i.e., m=0, and find the relation is inconsistent with this with a probability of p=\Prefl. 

Figure \ref{fig:data}b shows a positive correlation between the corona-disk path length, derived from reverberation time lags, and the radio Eddington luminosity.  A linear fit in log space to the {\it XMM-Newotn} data is given by $\log H = $\mH$ \log (L_R/L_{Edd}) +$\bH with an intrinsic scatter of $\sigma_H=$\sigH (Table \ref{Tab:modelparams}). Again, we utilized an MCMC model, drawing the data from a Normal distribution using 1-$\sigma$ uncertainties and a uniform distribution between zero and 3$\sigma$ upper limits in linear space for the non-detections. This fit is also inconsistent with the null hypothesis of no correlation, i.e. m=0, at $p=$\PH . In addition, this fit is inconsistent with a slope of unity at $p<$\PHone . This is important to note as both the path length and radio Eddington luminosity scale inversely with mass, by definition. Therefore, if mass was singularly driving this relation, the slope would be consistent with unity, which we do not find. 

Together, these two relations (Figures \ref{fig:data}a\& b) indicate that changes in the corona are linked to jet production in AGN.

\subsection{The Corona-Jet Model}

We next investigate a physically motivated model that predicts both changes in reflection fraction and disk-corona distance as a function mass-accretion rate and, consequently, jet power. These changes are assumed to originate in a corona that is moving away from the disk. As the velocity increases, more of the coronal radiation is beamed away from the accretion disk, decreasing the reflection fraction. If the corona is outflowing from the disk, it is reasonable to assume that it is associated with the base of the jet, and that increases in velocity in the corona propagate into the jet. The increased velocity in the jet results in more efficient synchrotron radiation, and thus a more radiatively powerful jet.  Finally, an increase in mass-accretion rate is assumed to drive the increase in velocity. The higher mass-accretion rate provides both a larger radiation field to push on the charged particles via Compton scattering, and plausibly a stronger magnetic field, which can dissipate in the corona, also accelerating the charged particles \citep[e.g.][]{Kara16}. In addition, the local sound speed may also set the corona velocity \citep{Markoff05}, which increases with mass-accretion rate. 

\begin{figure*}[thp!]
\center
\subfigure[][]{\includegraphics[width=.45\linewidth]{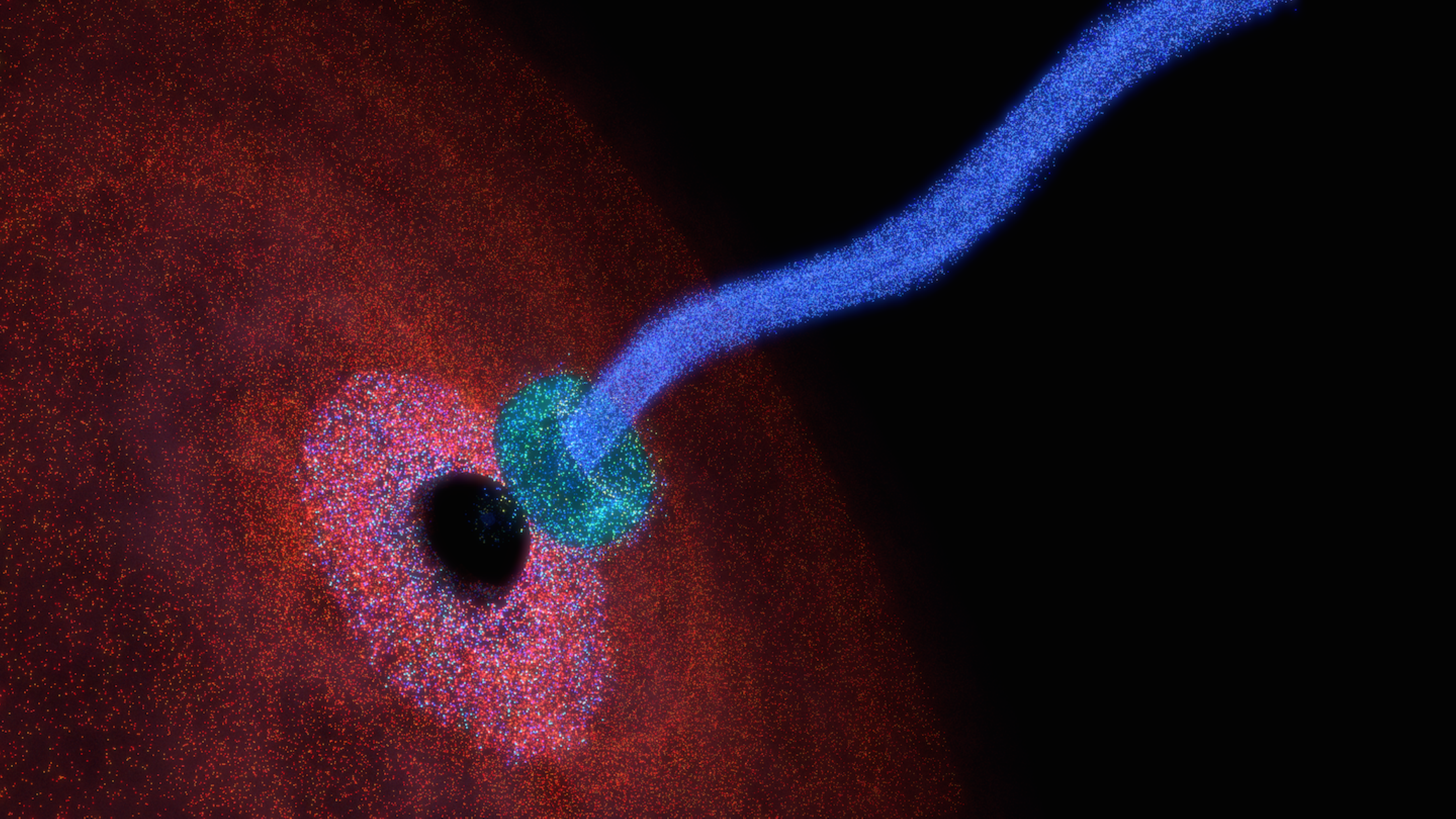}}
\subfigure[][]{\includegraphics[width=.45\linewidth]{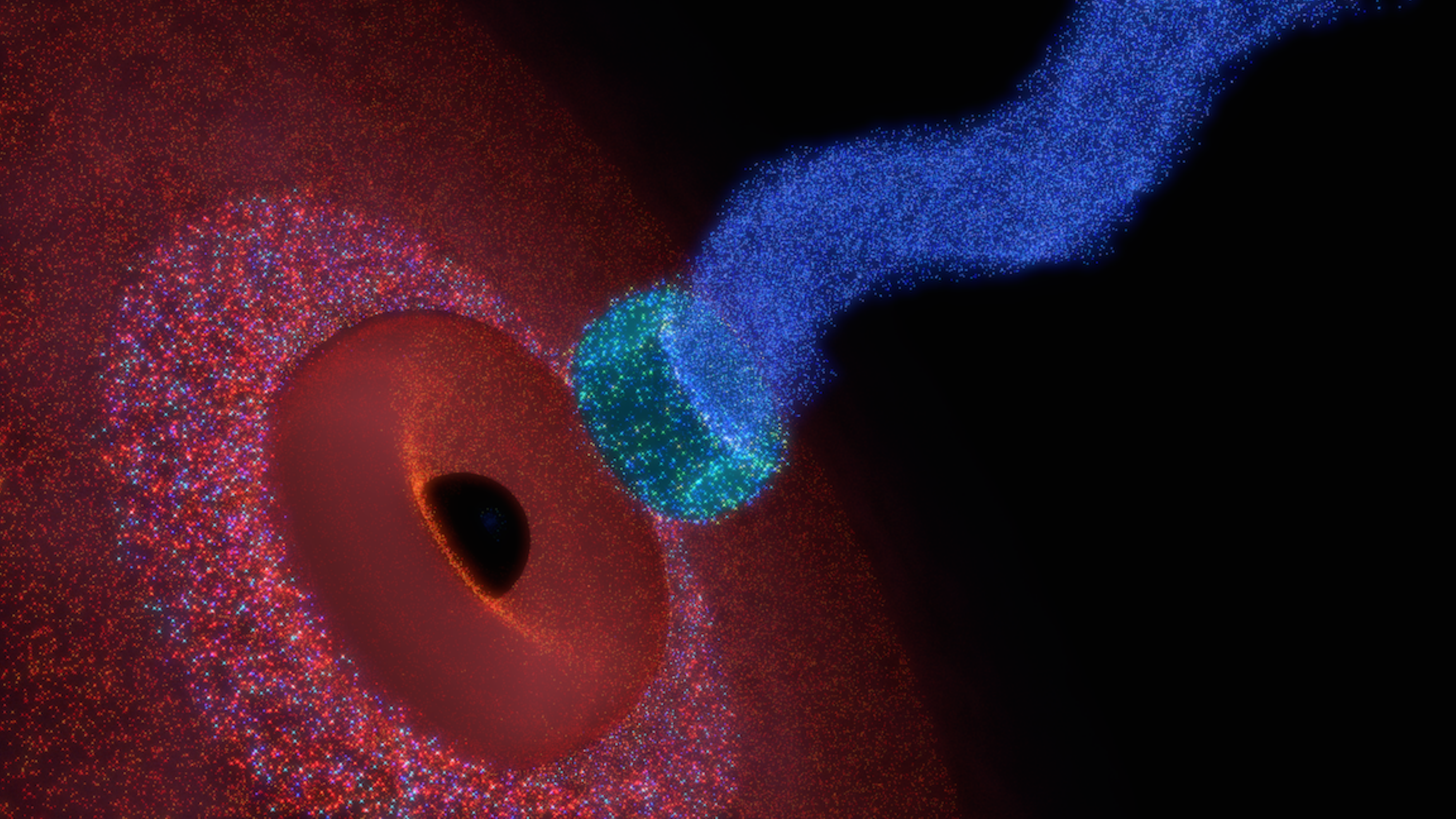}}
\caption{This cartoon depicts our model that suggests the reflection region and jet power increases as a black hole moves from a low accretion rate (a) to a higher accretion rate (b). The accretion disk is denoted with red particles while the corona and jet are denoted by green and blue particles, respectively. The corona shines down on the disk, causing the inner regions to fluoresce, denoted by the brightest regions in the disk surrounding the black hole. As the mass-accretion rate increases from (a) to (b), a region of the inner disk becomes radiation pressure dominated and vertically extended (torus-like structure in b). Eventually the ionization in this region becomes high enough that the reflection features are not easily detected, and the reflection is measured further out in the disk. In addition, as the mass-accretion rate increases, the corona moves faster and the jet increases in power, depicted in these figures as a wider extent of the jet. These figures are not to scale, and were created using the open source 3-D visualization software Blender (https://www.blender.org).   \label{fig:cartoon} }
\end{figure*}

In addition, as the mass-accretion rate increases, a region in the inner disk becomes radiation dominated. The ionization of this region increases and the reflection fraction decreases \citep{Garcia13}. At high enough mass-accretion rates, the inner region becomes so optically thick that the radiation is advected across the event horizon, again dramatically decreases the reflection features \citep{Narayan98}. Therefore, the most prominent ``reflection'' region moves outward in the disk to the gas-pressure dominated regions, increasing the observed time-lags, and therefore the predicted  path length between the corona and emission region in the disk (Figures \ref{fig:cartoon}a\&b).  In the following subsections each of the particular model components (moving corona, jet, and disk) are described in detail.

\subsubsection{The Moving Corona}
\cite{Beloborodov99} describe a model for computing reflection fractions for a moving corona. They developed this model to fit the small reflection fraction in stellar-mass black hole Cygnus X-1. As a result, the model not only fit the small reflection fraction in Cygnus X-1 ($R\sim0.3$) but also the hard spectral index of $\Gamma=1.6$ with a moving corona with velocity $\beta=v/c=0.3$. 

In this model, as the corona moves away from the disk, the photons get Doppler shifted away, resulting in a decrease of flux illuminating the disk. This results in a reflection fraction that is dependent on velocity as:
\begin{equation}
R = \frac{ (1+\beta/2)(1-\beta\mu)^3}{(1+\beta)^2}
\end{equation}
where $\mu=\cos \theta_\mu$,  $\theta_\mu$ is the viewing angle to the source.

\subsubsection{Self-Similar Jet}
In order to compare the predicted reflection fraction to the jet luminosity, we next assume the radio luminosity is well described by the self-similar, optically thick jet model. Synchrotron emission from a self-similar jet that is optically thick, $\alpha=0$, where $S_\nu\propto \nu^\alpha$,  scales with both black hole mass, $M$, and mass-accretion rate, $\dot{m}=\dot{M}/\dot{M}_{Edd}$ as, 
\begin{equation}
L_R \propto M^{187/120} \dot{m}^{17/15}.
\end{equation}
where $L_R$ is the radio luminosity, and we assume a standard thin, gas-pressure dominated disk \citep{Heinz03}. The magnetic field in a gas-pressure dominated disk also scales with black hole mass and mass-accretion rate as 
\begin{equation}
B^2 \propto \dot{m}^{4/5} M^{-9/10}
\end{equation}
\citep{Heinz03}. Finally, we assume that the gas pressure in the jet is proportional to the magnetic field strength,
\begin{equation}
B^2 \propto \rho_j(\gamma-1)c^2
\end{equation}
 where $\gamma=(1-\beta^2)^{-1/2}$, and $\rho_j$ is the gas density in the jet. We note that we keep $\rho_j$ fixed, but this could likely change as a function of mass-accretion rate, mass of the black hole, and even magnetic field strength, depending on how the jets are loaded.
 
Combing Equations 2, 3 \& 4 gives the following radio luminosity dependence on velocity: 
\begin{equation}
\frac{L_R}{L_{Edd}} = N M (\gamma-1)^{17/12}
\end{equation}
where where $L_{Edd}=1.38\times10^{38} (M)$ ergs $s^{-1}$, and $N$ is a normalization constant that we fit for in our final MCMC simulations.

\begin{figure*}[!thp]
\center
\subfigure[][]{\includegraphics[width=.45\linewidth]{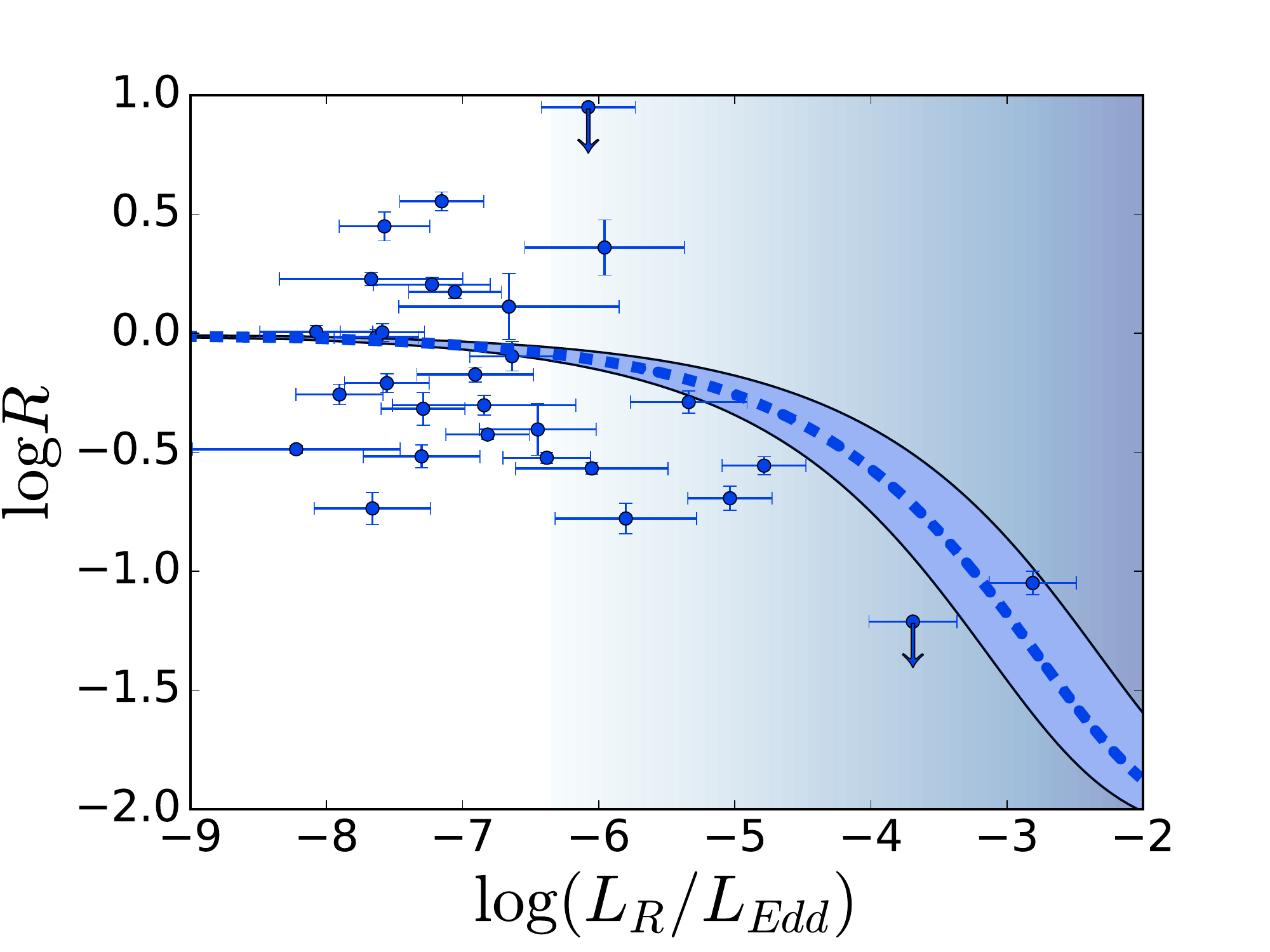}}
\subfigure[][]{\includegraphics[width=.45\linewidth]{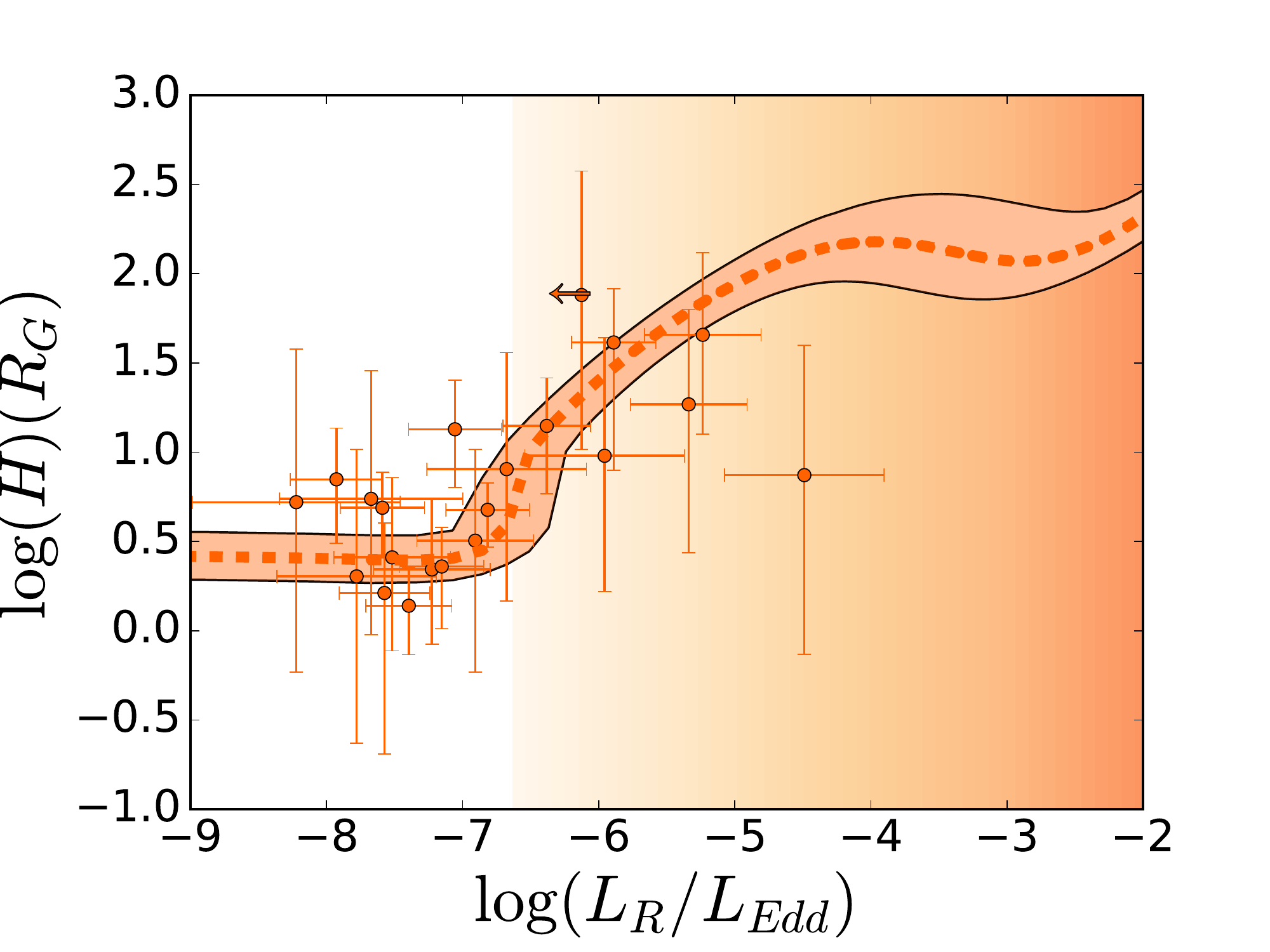}}
\caption{This figures shows the best fitting model of a moving corona and self-similar jet emission overlaid on our data (Table \ref{Tab:modelparams}). We assume an average mass of $\langle \log M\rangle=7.8$ and $\langle \log M\rangle = 7.1$ for the model displayed in panels a) and b) respectively, and the solid shaded regions give the resulting $1\sigma$ confidence levels in the vertical direction. The gradient shaded regions denote the AGN that have radiation pressure dominated inner disks. The darker the shading the larger the radiation-dominated region. \label{fig:bestfit} }
\end{figure*}

\begin{figure*}[thp!]
\center
\subfigure[][]{\includegraphics[width=.45\linewidth]{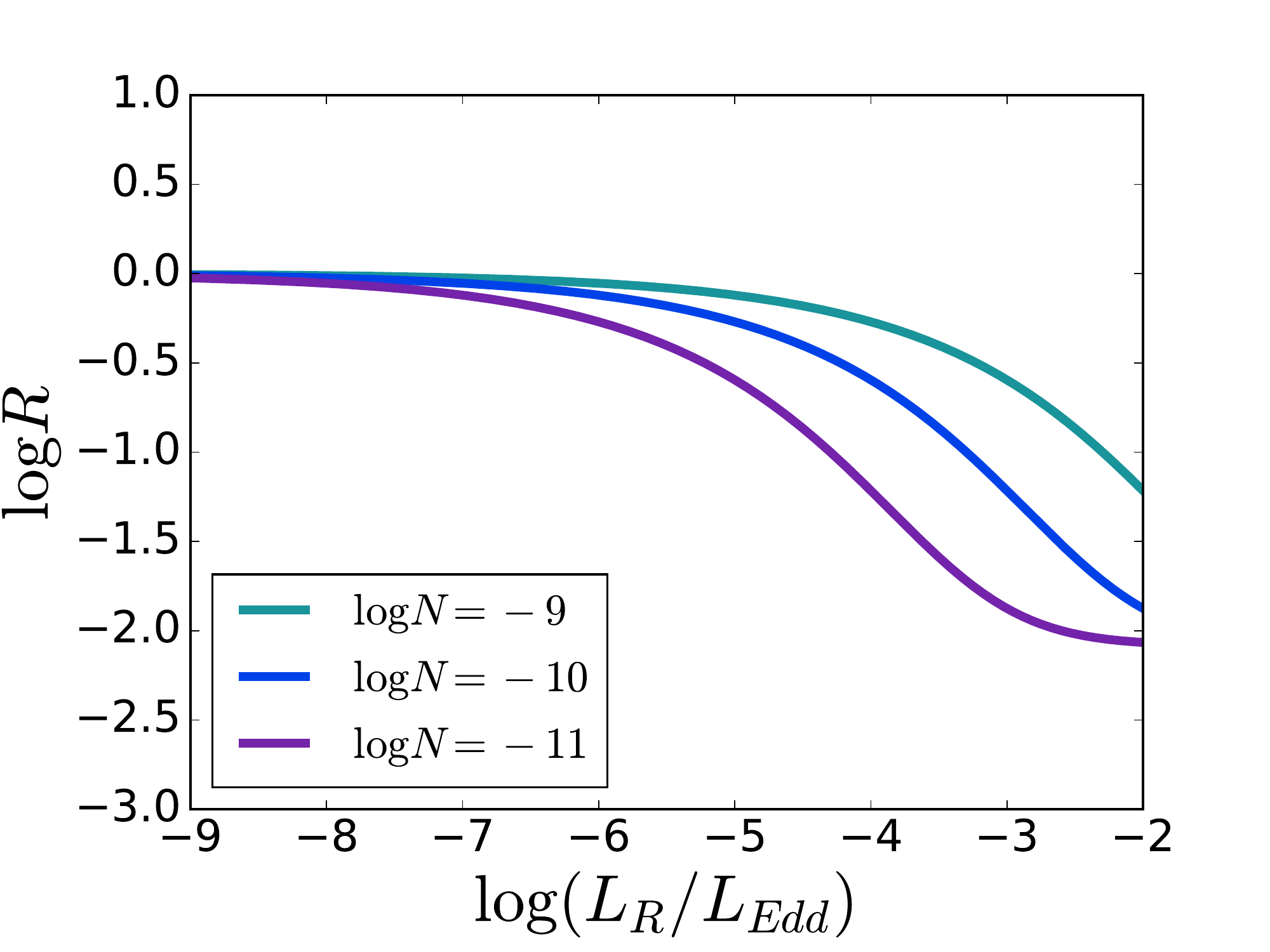}}
\subfigure[][]{\includegraphics[width=.45\linewidth]{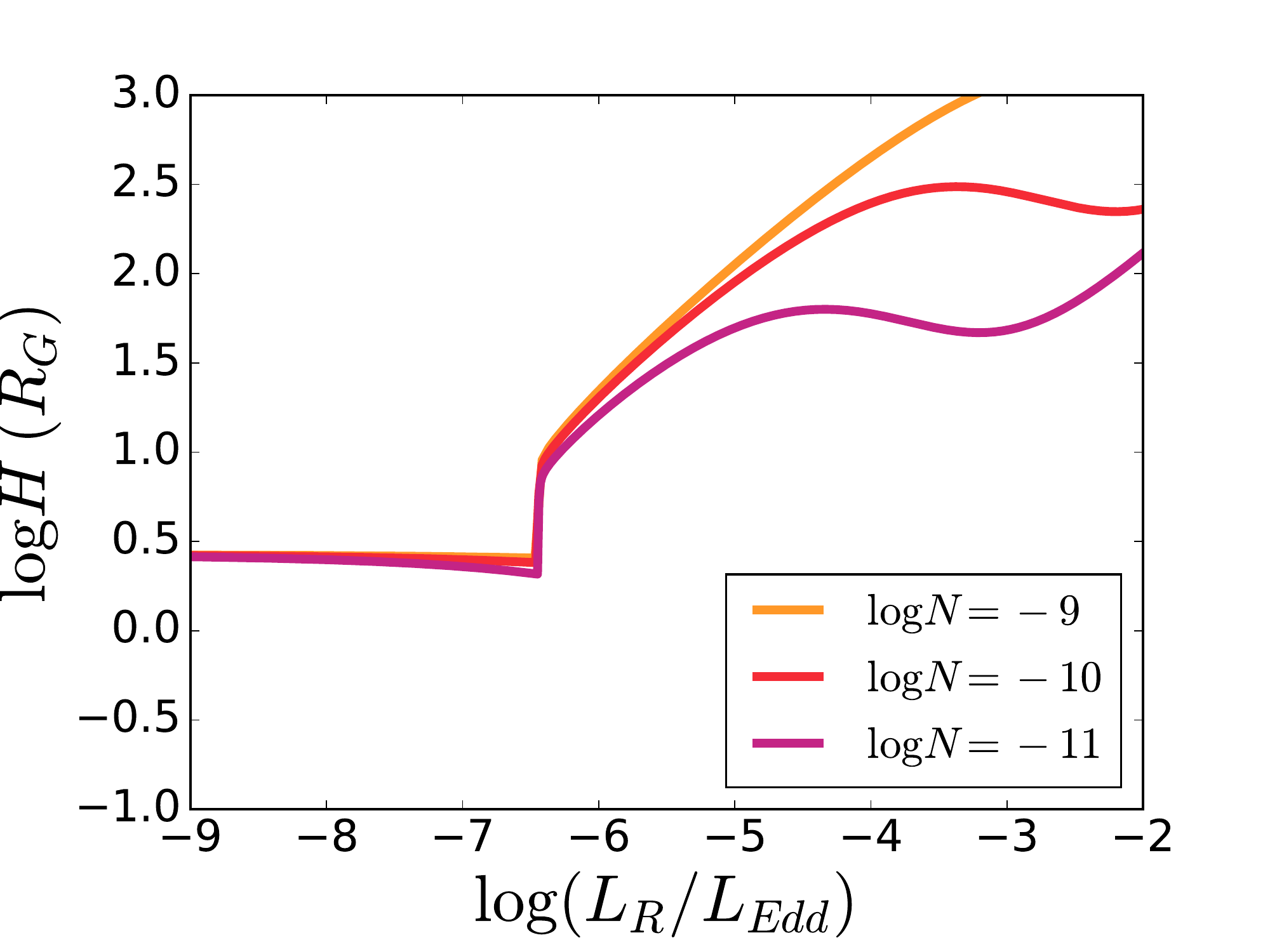}}

\subfigure[][]{\includegraphics[width=.45\linewidth]{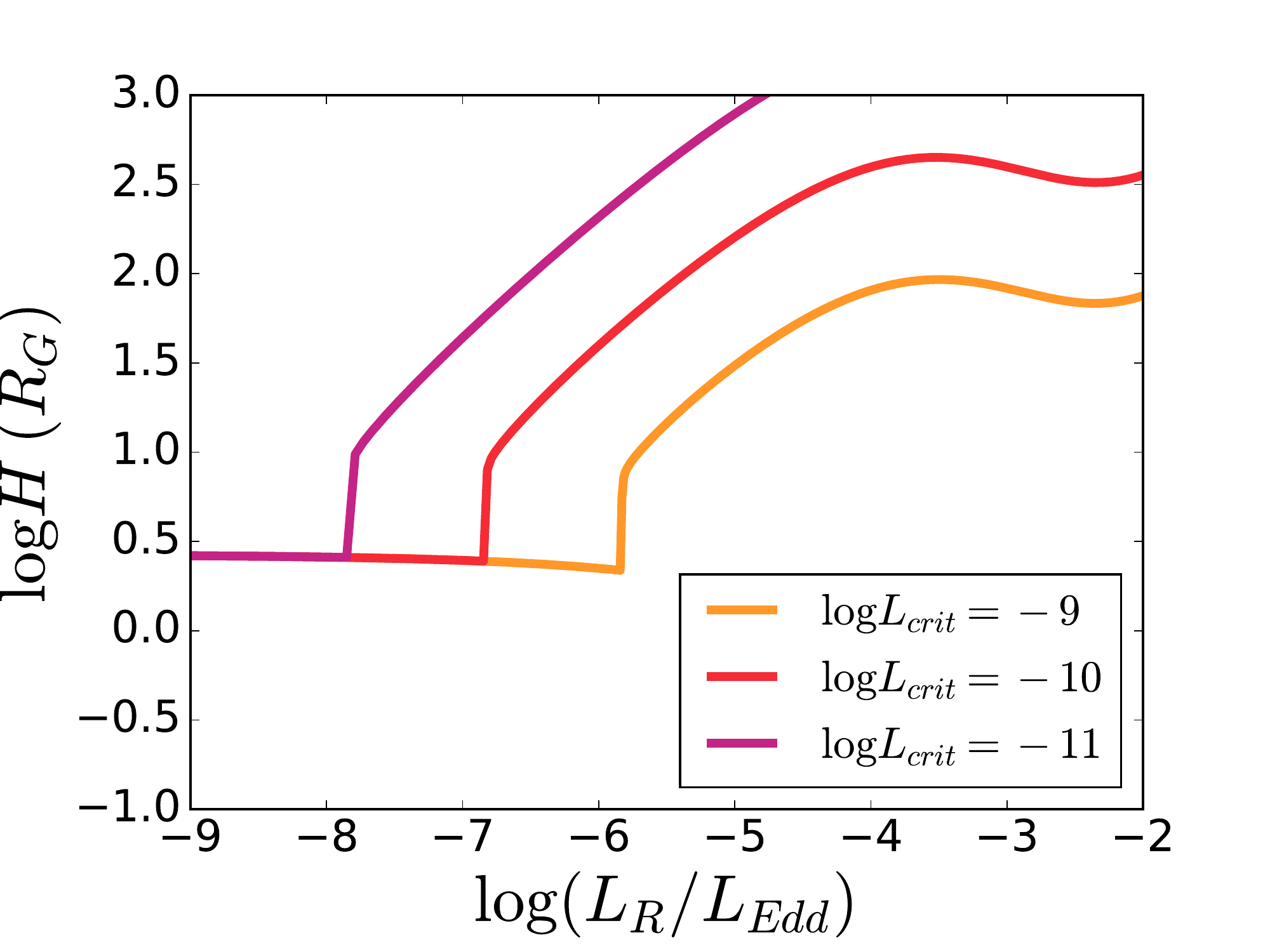}}
\subfigure[][]{\includegraphics[width=.45\linewidth]{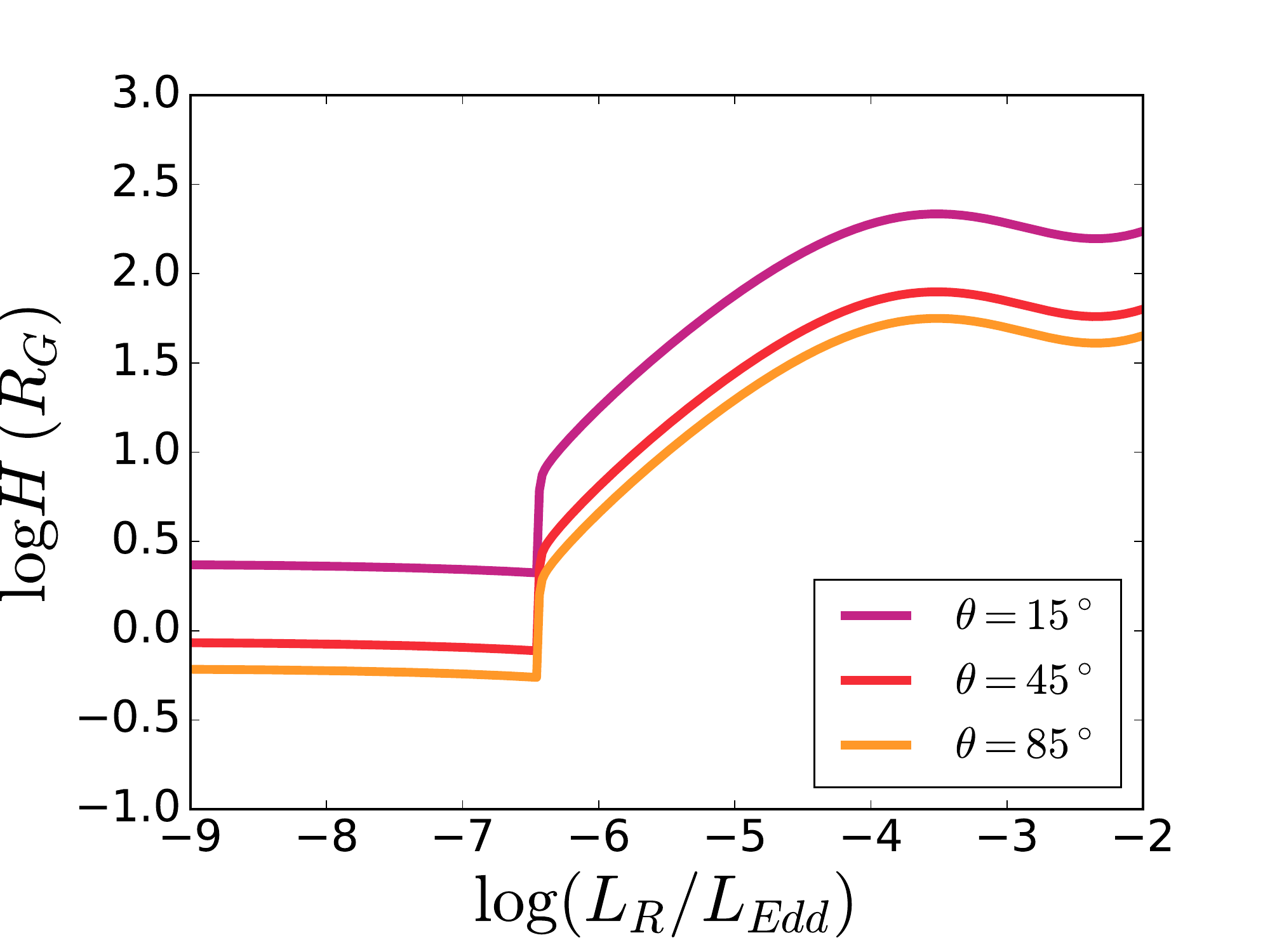}}

\subfigure[][]{\includegraphics[width=.45\linewidth]{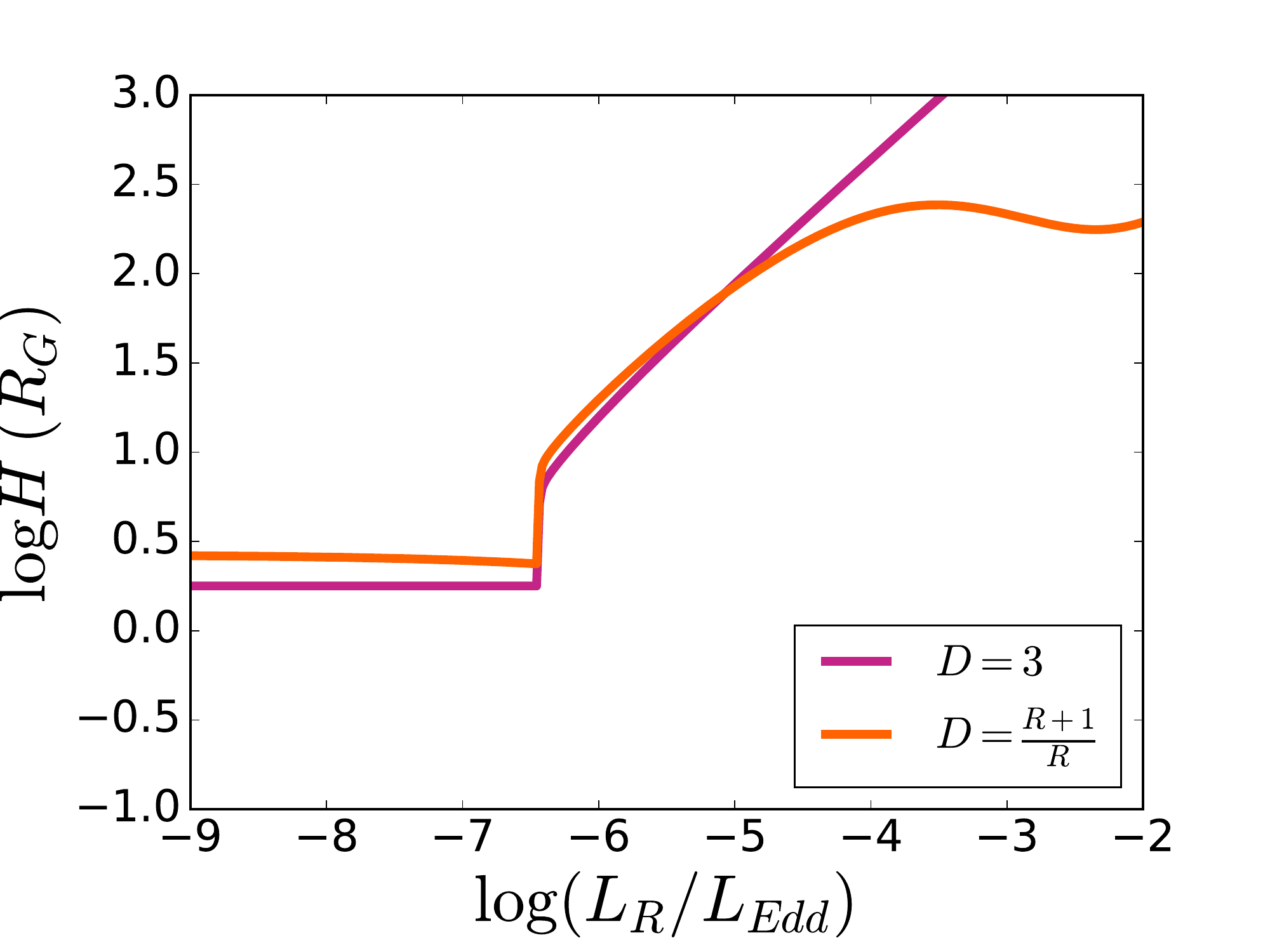}}
\caption{ This figure shows how varying the different fit parameters, $\log N$, $\log L_{crit}$, $\theta$, affects the best fit model shown in Figures \ref{fig:bestfit}a\&b. The varied parameters are shown in the legend of each subplot with all other parameters fixed to the best fit values in Table \ref{Tab:modelparams}.  Panel e) shows the effects of including dilution as a constant versus depending on the reflection fraction as suggested by \cite{Cackett14}. \label{fig:modelparams} }
\end{figure*}

\subsubsection{Standard Accretion Disk}
In addition to predicting the reflection fraction from a  moving corona and the luminosity from the optically thick jet , we can use the model to predict the relative path length between the corona and the disk, $H$. As $H$ is estimated from the X-ray reverberation time-lags, we assume it is the shortest path between the corona and the most efficient reflecting region in the disk, $R_{refl}$. $R_{refl}$ is typically assumed to be the inner radius of a thin disk, as this is the region that has the highest emissivity to produce both the Fe K$\alpha$ line and Compton hump reflection features \citep{Reynolds99}.  However, high mass-accretion rates can result in a region in the inner disk that is radiation pressure dominated. The inner layers in this region may become overly ionized as it increases in scale height and temperature, thus unable to efficiently produce distinct reflection features \citep{Shakura73,Garcia13}. The boundary, $r_{b}$, between the radiation pressure and the standard, gas-pressure dominated regions scales as:

\begin{equation}
\begin{cases}
\frac{r_b}{(1-r_b^{-1/2})^{16/21}} = 150(\alpha_D M)^{2/21} \dot{m}^{16/21}, & \dot{m} \gtrsim \frac{1}{170} (\alpha_D M)^{-1/8} \\
r_b = r_{isco}, &\dot{m} < \frac{1}{170} (\alpha_D M)^{-1/8}
\end{cases}
\end{equation}
where $\alpha_D$ parameterizes the viscosity in the disk, and $r_{isco}$ is the innermost stable circular orbit (ISCO) \citep{Shakura73}. Combining Equations 2 and 6 and assuming $R_{refl}=r_b$,  we find the inner disk radius, where the reflection features can easily arise, scales with luminosity as:

\begin{equation}
\begin{cases}
 \frac{R_{refl}}{(1-R_{refl}^{-1/2})^{16/21}} =   N_2 \left( \frac{L_R/L_{Edd}}{M^{5/12}}\right)^{80/119}, & \frac{L_R}{L_{Edd}} >  L_{crit} {M}^{5/12}  \\
R_{refl} = R_{refl,0}, & \frac{L_R}{L_{Edd}} < L_{crit} {M}^{5/12} 
\end{cases}
\end{equation}
$L_{crit}$ is a combination of the factor that scales the mass and mass-accretion rate to an observable 1.4 GHz luminosity ($L_R$, Equation 2) and the normalization, $\frac{1}{170\alpha_D^{1/8}}$, in the conditional dependence of Equation 6.  We assume $R_{refl,0}=1.23697 R_G$, i.e. the smallest inner radius for a maximally spinning black hole, $a=0.998$ \citep{Bardeen72}. Making sure the distribution is continuous, we re-write $N_2$ as: 
\begin{equation}
N_2 = \frac{R_{refl,0}}{(1-R_{refl,0}^{-1/2})^{16/21}} L_{crit}^{-80/119}.
\end{equation}
Interestingly, the reflection fraction is only sensitive to changes in the scale height at very small radii, $<10 R_G$ \citep{Dauser16}. Therefore, for the simplicity of the model, we ignore this effect for the reflection calculations and only include them in the coronal height calculations. We also stress the assumption of $R_{refl}=r_b$ is an over simplification, and that there likely is still some emission coming from the radiation pressure dominated regions, especially during the initial transitions.  However, it is still illustrative in this first-order approximation we are making (See Section 4.3 for further discussion).

Finally, we take into account the dilution of the time lags due to varying contributions of the reflected spectrum as a function of energy. Dilution is correlated with the percentage of observed reflected emission as compared to the continuum emission, and scales as $D=\frac{R+1}{R}$ \citep[see][for more details]{Cackett13,Cackett14,Wilkins13}. Figure \ref{fig:modelparams}e shows the effects of including the dilution factor in our model by plotting the corrected path length as a function of reflection fraction. The dilution factor as a function of reflection factor decreases the time lag and therefore inferred path length, $H$:

\begin{equation}
H  = \frac{R_{refl}}{ D\sin\theta}.
\end{equation}
where $\sin\theta$  is assumed to be the same for the entire sample.

\subsubsection{MCMC Setup}
We jointly fit the {\it NuSTAR} reflection strength sample and {\it XMM-Newton} coronal time-lag sample as functions of radio Eddington luminosity with the moving corona, self-similar jet, and standard thin disk accretion model using Equations 2, 5 \& 10 and an MCMC. The only free parameters in this model are the geometry, $\theta$, the normalization, $N$, and the critical luminosity, $L_{crit}$.  Because the dilution factor depends on reflection fraction, the normalization constrains both the  {\it NuSTAR} and {\it XMM-Newton} distributions. This can be seen in Figures \ref{fig:modelparams}a\&b, where we hold all other parameters constant and change the normalization.  Figures \ref{fig:modelparams}c\&d show the results of holding all other parameters constant and changing the critical luminosity or viewing angle, respectively. One can see that these later parameters only effect the {\it XMM-Newton} time-lag samples. 

We assume a log-normal prior on all the detected data, and a uniform prior between 0 and the $3\sigma$ upper limits in linear space for the non-detections. We assumed a log-uniform prior for the critical luminosity, $-13<\log L_{crit}<-6$, as well as uniform in the normalization, $-15<\log N<-7$.  Finally, we assume a uniform prior on $0 <\sin \theta<1$. We used a total of 100,000 steps, burning the first third of the distribution for each MCMC run.  After checking for convergence using four independent runs utilizing a Gelman-Rubin statistic for each parameter in question \citep{Gelman92}, the final results are given in Table \ref{Tab:modelparams} and over plotted in Figures \ref{fig:bestfit}a\&b. 

\subsubsection{MCMC Results}
The results of the MCMC fits are given in Table \ref{Tab:modelparams} and plotted in Figures \ref{fig:bestfit} a\&b. We find that our model does fit the data well, and has scatters very comparable to the individual linear fits. The small $\theta=$\jettheta indicates the path length inferred from the time lags between the corona and reflecting regions in the disk are on order of the corona height above the disk. For emission coming from close to the ISCO of a maximally spinning black hole at mass-accretion rates a few percent of Eddington, this corresponds to a height of $\sim 5 R_G$. This is consistent with heights measured in several radio-quiet Seyfert AGN where the disk extends down to the ISCO \citep{Miniutti04}. 

The critical luminosity is measured to be $\log L_{crit}=$\jetLcrit. This corresponds to a radio Eddingotn luminosity of $\log (L_R/L_{Edd})\approx -6.5$ using the average mass of the sample of $\langle \log M\rangle=  7.6$. In our model, this parameter determines at what radio Eddington luminosity the height of the corona is no longer constant but begins to scale with radio Eddington luminosity and mass of the black hole (Figure \ref{fig:modelparams}b). Physically, $\log L_{crit}$ determines when the inner disk structure becomes radiation pressure dominated \citep{Shakura73}. The shaded regions in Figures \ref{fig:bestfit} a\&b mark the region where the inner disk is radiation pressure dominated given the average mass of the distribution, with the darker regions corresponding to larger regions of the inner disk dominated by radiation pressure. 

Finally, we determine the normalization to be $\log N =$\jetN. The normalization is responsible for setting the position of the inflection point in Figure \ref{fig:modelparams}a as well as the amount of dilution which effects $H$ at high radio Eddington fractions in Figure \ref{fig:modelparams}b. The normalization also sets the scale for the velocity of the corona. Taking the average mass of our samples, $\langle \log M\rangle=  7.6$, the predicted velocities of the corona span \jetV, across our radio Eddington luminosity range. 

\section{Discussion} 
\subsection{Reflection Fractions}
Our sample indicates an inverse trend between reflection fraction and Eddington scaled radio luminosity. This is well characterized by a linear fit in log space and our physically motivated, moving corona-jet model. In this study, we measure the total reflection fraction with a \texttt{xillver} model. This is likely to be the sum of the relativistically blurred reflection from the inner accretion disk and the distant reflection from the outer accretion disk or torus. Contrary to our interpretation of a continuous distribution of reflection fractions,  one could interpret this inverse correlation as a constant distant reflection fraction from a torus with a contribution of blurred reflection in only the low radio Eddington sources. A survey by \cite{Nandra07} hints that the distant reflection is constant as it shows less scatter compared to the blurred disk reflection. In addition, the authors find the average reflection strength in Seyferts with detected blurred disk lines dominates over the distant reflection at $\langle R_{disk}\rangle = 1.00\pm0.18$ compared to $\langle R_{dist} \rangle = 0.46\pm0.06$. Likewise, many studies of radio AGN indicate that the reflection quantified by the neutral Fe K$\alpha$ line is very narrow, possibly originating from the disk \citep{Lee02} but likely originating in a torus \citep{Nandra94}.

Assuming the standard AGN unification model \citep{Krolik94}, where AGN have the same structure with viewing angle being responsible for the difference in observed spectra, we could expect that all AGN have a similar distant reflection fraction from the torus. Therefore, we might expect a ``floor'' in the reflection fraction. Our highest radio Eddington sources strongly argue against this, as they are well below the average reflection fraction for the distant reflectors in Seyfert galaxies given by \cite{Nandra07}. This suggests that either 1) the distant torus is nearly absent in the high radio luminosity sources, contrary to the standard unification model, or 2) that the distant reflecting region is also sensitive to the beaming of the corona. 

There is strong evidence that 3C 273, our highest radio Eddington luminosity source, still has a substantial torus as supported by detections of molecular hydrogen in the vicinity of the Quasar \citep{Kawara89}.  A followup study by the authors indicates that the mass in Quasar tori (3C 273 and IRAS 13349+2438) is roughly six times larger than in Seyfert galaxies \citep{Kawara90}, strongly disfavoring the idea that the torus is diminishing as AGN move to higher mass-accretion rates. Therefore, we favor the interpretation that both the disk and distant reflection are sensitive to beaming of the corona. 

We also examine the hint of larger scatter at low radio Eddington fractions in Figures \ref{fig:data}a \& \ref{fig:bestfit}a. The largest outlier and highest reflection data point is PG 1247+267. At a redshift of $z=2.038$, PG 1247+267 is also an outlier in terms of distance as the rest of the sample resides below $z<0.2$. It therefore likely resides in a different environment then the majority of our sample, as the universe was much denser at $z=2.038$. This could have effects on both the structure of the disk and jet, and resulting in different measured reflection and jet properties, resulting in PG 1247+267 being an outlier from the sample distribution.

The rest of the scatter at low radio Eddington fractions may be due to varying modes of accretion. At a mass-accretion rate of a few percent of Eddington, the inner disk is predicted to move from a standard, optically thick, geometrically thin accretion disk to an advection dominated flow \citep{Narayan95,Narayan98}. In such a regime, the inner disk may no longer be as efficient at producing reflection features, and the jet luminosity scales differently with mass and mass-accretion rate as is assumed in Equation 2. Both of these would increase the scatter of our distribution. 

The scatter in reflection measurements may also be due not including relativistic effects as the corona recedes closer to the black hole, as is implied by Figure \ref{fig:data}b \& \ref{fig:bestfit}b. Our model currently saturates at a reflection fraction of unity because we assume that the velocity is always outflowing from the disk as defined by jet structure. If we allowed for velocities to approach the disk, we could of course get $R>1$ \citep{Reynolds97,Beloborodov99}, but would not be able to calculate the corresponding jet luminosity in the analytical form presented here. In the future, we aim to include general relativistic effects into our calculations of the reflection fraction with the goal of better characterizing the low radio Eddington fraction data. This regime is where the time-lag analysis data predicts the smallest coronal height above the disk (Figure \ref{fig:bestfit}b), i.e. the region most likely be effected by gravitational light bending and relativistic effects. Such an effect would boost the reflection fraction above unity \citep{Tanaka95,Fabian00}, in agreement with our highest reflection fraction measurements. 

\subsection{Corona-Jet Velocities}
In addition to quantifying how the AGN reflection fraction and the corona path length to the reflecting region of the disk scales with radio Eddington fraction, our model predicts the coronal and subsequent jet velocity also scales positively with radio Eddington fraction. 

Stochastic motions in the corona have been observed in individual AGN, including in Mrk 335 \citep{Parker14,Wilkins15} and 1H 0707-495 \citep{Fabian09,Fabian12,Wilkins14}. Through the use of detail modeling of the Fe K$\alpha$ line, the reflection fraction is inversely correlated with flux \citep[e.g.,][]{Zdziarski99,Parker14,Wilkins14,Wilkins15}. This is consistent with the height of the corona increasing above the disk reducing the amount of light bending, which ultimately decreases the reflection fraction, and increases the continuum flux. \cite{Wilkins16b} take this idea one step further and suggest the flaring activity in the low flux states may be reconfigurations of the corona into a base of a jet. Our own analysis suggests more dramatic changes in the corona during even larger changes in accretion states then examined in these works, but qualitatively agree with findings of a moving corona.

Physically explaining stochastic motions requires dissipation of energy into the corona and subsequently the jet. This motion of the corona has therefore been explained via advection dominated accretion disks creating outflows \citep[e.g., ][]{Blandford99,Quataert00}, Compton pressure from the disk and reflected photons on the corona \citep{Beloborodov99,Ghisellini10}, or magnetically driven corona above an accretion disks \citep{Merloni02,Hawley06,Tchekhovskoy16}. In particular, these last two are most likely applicable to our sample at intermediate to high mass-accretion rates. Interestingly, MHD models of an outflowing corona above an accretion disk  naturally suggest that the velocity increases with height above the accretion disk \citep{Merloni02} as well as mass-accretion rate \citep{Ghosh97}. This is consistent with our models, which assume the velocity scales proportionately to radio luminosity and consequently mass-accretion rate and height of the corona. Conversely, the Compton pressure from the reflected photons as suggested by \cite{Beloborodov99} is also a viable model, and occurs simultaneously. This effect is most efficient when the plasma is made up of electron-positron pairs. Several works suggest that electron-positron pair production can regulate the temperature of the corona \citep[e.g.,][]{Svensson84,Zdziarski85,Pietrini95,Stern95,Dove97,Coppi99}. Recent work by Fabian et al (2015) comparing theoretical predictions to {\it NuSTAR} observations seems to confirm this, giving viability to this Compton rocket-like scenario.

On much larger scales further down the jet, changes in velocity have been used to explain the differences in jet structure in Fanaroff-Riley (FR) I vs FR II jet morphologies. In our model, we assume the velocity of the corona is proportional to the velocity accelerated further down the jet and therefore is relevant to large scale jet velocities as well. FR I jets are radio-loud jets with low surface brightness and diffuse structures, while FR II are jets at even higher luminosities and are highly collimated, terminating in large impact lobes. A slower velocity in FR I's compared to FR II's would would explain the morphology as the jets would be less collimated and more likely to diffuse into the surrounding medium \citep{Gendre13}. Though it can not be ruled out that environment also plays a role in these jet morphologies, it is clear that FR I's accrete at a lower rate than FR II's \citep{Marchesini04,Best12}. Thus it would not be surprising if the flow onto the black hole is intimately connect to the large scale jets, as we also suggest here. 

In addition, \cite{King15} examine the velocity of jet knots close to the black hole using Very Long Baseline Array (VLBA) archival data, and find that the velocity of these knots increases with X-ray Eddington luminosity. whether the X-ray emission is from synchrotron emission from the base of the jet or Comptonized photons from the disk, the correlation indicates the velocity scales with increasing photon energy density close to the black hole, likely a result of increasing mass-accretion rate. These knot velocities are much larger than the predicted for the corona,  \jetV. This is understood in the context of these VLBA measured knots being measured on order of a  $10^4 R_G$ from the black hole, and have already undergone some acceleration along the jet.

\subsection{Radiation Pressure Dominated Inner Disk}
Along with mildly relativistic corona velocities, our model predicts that many of the sources are accreting close to their Eddington rates and enter a vertically extended, radiation pressure dominated accretion phase in regions closest to the black hole (Shaded region in Figures \ref{fig:bestfit}a\&b). Our model assumes that no reflection features are emitted in this radiation pressure dominated regime due to the gas being overly ionized. In reality, the ionization is likely to increase gradually, being dependent on both mass and mass-accretion rate. Thus, the reflection should still originate close to the ISCO in several of the sources above $L_{crit}$ (lightly shaded region in Figures \ref{fig:bestfit}a\&b). This would reconcile many of the analyses of AGN, like IRAS 13224-3809 \citep{Fabian13}, that have detailed reflection measurements emitting from the ISCO but where we predict a radiation pressure dominated inner disks. 

The exact transition of when the inner disk becomes overly ionized is out of the scope of this paper. Our current model does not allow us to constrain this due to our assumption of ``total'' ionization as soon as the region becomes radiation pressure dominated. Furthermore, the value of $L_{crit}$ is ultimately constrained by our assumptions of how the coronal velocity and height scales with mass-accretion rate and mass of the black hole. In the future and with more data, we plan to allow these relations to vary freely to more realistically characterize $L_{crit}$. 

Though a rough estimate, $L_{crit}$ is still corroborated by observations of radio-loud AGN that indicate the reflection region is moving outward. High resolution observations utilizing {\it XMM-Newton} show that, in addition to a narrow line component, 4C 74.26 has a broad Fe K$\alpha$ line coming from the inner disk but located at a distance $>50R_G$ \citep{Larsson08}. Such a large distance indicates that not all broad lines are emanating from regions exactly at the ISCO, possibly caused by over ionization of the inner disk. Therefore, we postulate that there is a continuum of reflecting regions along the disk at the highest mass-accretion rates. 

The radiation pressure at the highest mass-accretion rates may also play a large role in jet production. As the scale height of the disk increases as it becomes radiation pressure dominated, the resulting thick inner disk could be necessary for efficiently confining, collimating and even accelerating the natal corona into highly relativistic jets \citep{Sarazin80,Krolik07}. Simulations of jets generally support this idea, as they require a ``thick disk'' to produce jets and anchor sufficiently strong magnetic fields \citep[e.g.,][]{Meier01,McKinney12,Foucart16}.  Using the average mass of our sample, $\langle \log M\rangle=  7.6$, we put the critical luminosity, $\log L_{crit}=$\jetLcrit,  in terms of an observable radio luminosity of $\log L_{1.4GHz,crit} \approx \jetL + (17/12 \log M_{7.6})$ ergs s$^{-1}$. Comparing this value to the divide between ``radio-loud'' and ``radio-quiet'' jets, we find the critical luminosity is just below the divide at $\log L_{5 GHz} \approx 41$ ergs s$^{-1}$ \citep{Miller90}. This indicates the inner disk at the divide is well into the radiation pressure dominated regime and may be necessary for launching the strongest jets. We predict the region dominated by radiation pressure at the dichotomy divide extends to $\sim \jetDichRad R_G$, utilizing Equation 7, which is consistent with the radius measured in the radio-loud AGN, 4C 74.26 \citep{Larsson08}. 

Finally, we note that the critical luminosity is also dependent on the mass of the black hole, which is consistent with studies of the radio-loudness dichotomy that demonstrate the importance of including mass as an additional parameter in the characterization of radio-loudness \citep{Broderick11}. \cite{Broderick11} show that the radio-loudness distribution is not actually a dichotomy but rather a continuum when mass is properly accounted for.  This agrees well with our model, as the inner disk gradually transitions to a radiation pressure dominated regime, and thus would gradually be more efficient at producing stronger jets.

\begin{deluxetable*}{lllllllll} 
\tabletypesize{\scriptsize}
\tablecolumns{9} 
\tablewidth{0pc} 
\tablecaption{Fit Parameters} 
\tablehead{ 
\colhead{Name}  & \colhead{slope} & \colhead{intercept} & \colhead{$\log(N)$} & \colhead{$\theta$} & \colhead{$\log L_{crit}$ } &\colhead{$\sigma_R$} & \colhead{$\sigma_H$}}  \\
\startdata
Log-Linear fits  \\ \hline
$\log R$ vs $\log L_R/L_{Edd}$ &  \mR & \bR & & && \sigR&  \\
$\log H$ vs $\log L_R/L_{Edd}$ & \mH & \bH & && & & \sigH \\
\\ \hline 
\\
Moving Corona & & &\jetN &\jettheta &  \jetLcrit &\jetsigR & \jetsigH \\

\enddata
\tablecomments{ This table shows the best fit values for our log-linear fits as well as with our corona-jet model. The last two column gives the intrinsic scatter in each data set. The models are plotted in Figures \ref{fig:data} \& \ref{fig:bestfit}, and errors are 1$\sigma$. \label{Tab:modelparams} }
\end{deluxetable*}

\section{Conclusion}

In this paper, we compare both reflection fraction and the corona height to jet production, as traced by the radio Eddington luminosity. We find an inverse correlation between reflection fraction and radio Eddington luminosity, and a positive correlation of the path length connecting the corona and reflecting regions of the disk to the radio Eddington luminosity. These correlations can be explained via a moving corona that is propagating into the large scale jets. Our corona-jet model does well to characterize the data, determining 1) the corona velocities are mildly relativistic, 2) the corona is ``far'' from the disk such that the path length measured from the time-lags are close to the corona's height above the disk, and 3) the inner disk becomes radiation pressure dominated just before the AGN become ``radio-loud''. The last point suggests that a thick inner disk, due to the large scale height of a radiation pressure dominated region, is needed to efficiently collimate and accelerate jets in AGN. 

Future X-ray observations will need to focus on the strongest radio jet sources. Though the predicted reflection fractions are small, and the  time-lags are large, it is imperative to increase our sample at the highest radio luminosities. This is the regime that constrains both the geometry of the corona and the disk. Both of these quantities are crucial to understanding outflows in these accreting systems. 

\begin{acknowledgements}
The authors would like to thank the anonymous referee for their insightful and helpful comments. The authors would also like to thank Dan Wilkins for productive discussions regarding this manuscript. ALK would like to thank the support provided by NASA through Einstein Postdoctoral Fellowship grant number PF4-150125 awarded by the Chandra X-ray Center, operated by the Smithsonian Astrophysical Observatory for NASA under contract NAS8-03060. AL acknowledges support from the ERC Advanced Grant FEEDBACK. EK would like to thank the Hubble Fellowship program. Support for program number HST- HF2-51360.001-A was provided by NASA through a Hubble Fellowship grant from the Space Telescope Science Institute, which is operated by the Association of Universities for Research in Astronomy, Incorporated, under NASA contract NAS5-26555.
\end{acknowledgements}

\bibliography{bib}
\end{document}